\documentclass{aa}
\usepackage{graphicx,epsfig}
\usepackage{amssymb,amsmath}
\usepackage{natbib}
\usepackage{aas_macros}
\include{def}
\bibpunct{(}{)}{;}{a}{}{,}
\newcommand{\etal}{{et al.\ }}
\newcommand{\lta}{\stackrel{<}{\scriptstyle\sim}}
\newcommand{\gta}{\stackrel{>}{\scriptstyle\sim}}

\begin{document}

\title{Wavelength and Redshift Dependence of Bulge/Total Light Ratios in Galaxies} 
\subtitle{}
\author{Jochen Schulz, Uta Fritze -- v. Alvensleben, Klaus J. Fricke}
\institute{Universit\"ats-Sternwarte, Geismarlandstr.~11, 37083 G\"ottingen, Germany}


\authorrunning{Schulz \etal}
\titlerunning{Bulge/Disk Light Ratios}
\offprints{U. Fritze -- v. A., \email{ufritze@uni-sw.gwdg.de}}
\date{Received xxx / Accepted xxx}

\abstract{HST has opened the possibility to decompose the surface brightness profiles of
galaxies up to significant redshifts and look-back times into ${\rm r^{1/4}
-}$ bulge and exponential disk components. This should allow to study the
redshift evolution of bulge and disk luminosity contributions and discriminate
between the different formation scenarios for these galaxy components
currently discussed, i.e. decide if star formation in bulges and disks started
at the same time or was delayed in either of the two components. An
indispensable prerequisite for the comparison of bulge-to-disk ratios of
galaxies at different redshifts is to properly account for cosmological band
shift and evolutionary effects.\\
We present evolutionary synthesis models for both components and add their
spectra in various proportions to obtain the full range of local galaxies'
B-band bulge-to-total light ratios. Bulge star formation is assumed to occur on
a short timescale of $10^9$ yr, disk star formation proceeds at a constant rate.
We study the evolution of the relative light contributions of both
components backward in time and, for a given cosmological model, as a function
of redshift. This allows us to see how far back into the past the locally
well-established correlation between galaxy morphologies and spectral properties
can hold. To cope with the present uncertainty about the formation epochs of
bulge and disk components we present models for three scenarios: bulges and
disks of equal age, old bulges and delayed disk star formation, and old disks
with subsequent bulge star formation.\\
We quantitatively show the wavelength dependence of bulge-to-total ($=$ B/T)
light ratios for local galaxies. The different star formation timescales for
bulge and disk components lead to B/T ratios that significantly increase from U
through I-bands (by factors 4 -- 6 for weak bulge systems $\sim$ Sc) with the
rate of increase slightly depending on the relative ages of the two components.\\
The redshift evolution of B/T-ratios in various bands U, B, V, I, H is 
calculated
accounting both for cosmological and evolutionary corrections assuming a
standard cosmology (${\rm H_0 = 65,~\Omega_0 = 0.1,~\Lambda_0 = 0}$). In
particular, for the two scenarios with old bulges and old or younger disks, the
redshift evolution of B/T-ratios is dramatic in every band and both for galaxies
ending up at ${\rm z \sim 0}$ with low and high B-band B/T light ratios. Our
results clearly show that it does not make any sense to compare B/T ratios
measured in one and the same band for galaxies at different redshifts without
fully accounting for evolutionary and cosmological effects. These,
unfortunately, significantly depend on the relative ages of the two components
and, hence, on the galaxy formation scenario adopted. 
We also show that simultaneous decomposition of galaxy profiles in several bands
can give direct information about these relative ages and constrain formation
scenarios for the different galaxy components.\\
Of the wavelength bands we explore (U, B, V, I, H), the
I- and H-bands show the smoothest redshift evolution and, hence, are best
suited for a first order comparison of galaxies over the redshift
range from ${\rm z=0}$ to ${\rm z \gta 1}$. Our robust result that -- 
irrespective
of the respective ages of the bulge and disk stellar components --
I-band B/T-ratios apparently increase with increasing redshift for all
galaxy types with present B/T $> 0.1$ implies that the scarcity of
bulge-strong systems at ${\rm z \geq 0.8}$ reported by \citet{mar} and \citet{ag2002} for HDF and Hawaiian 
Deep Field galaxies is further enhanced.
}

\maketitle


%

\section{Introduction}

The bulge-to-disk ratio was one of the main qualitative classification criteria
that led \citet{1926ApJ....64..321H}
 to establish his famous Hubble sequence of morphological
galaxy types. When \citet{1958ApJ...128..465D} showed that surface
brightness profiles of M31 and other spirals could be decomposed into spheroidal
components following an ${\rm r^{1/4}}$ law and exponential disk components, a
quantitative decomposition became possible. \citet{simien} present bulge-to-disk
light ratios for 98 galaxies and investigate their relation to other galaxy
parameters, as e.g. the Hubble type.

Conventionally, Bulge/Disk {\bf(B/D)} or Bulge/Total {\bf(B/T)} light ratios for
local galaxies are measured in the B-band. Different Hubble types of galaxies
are mainly characterised by their respective B/T ratios.

Deep HST observations open up the possibility to determine structural parameters
of galaxies to considerable redshifts and to directly assess the interesting
question how galaxy morphologies have evolved cosmologically (cf.
\citeauthor{mar} \citeyear{mar}, \citeauthor{simard02} \citeyear{simard02}). 
\footnote{While it has been shown by \citep{andre1995} that bulges are more appropriately described by Sersic profiles ${\rm r^{1/n}-}$ profiles with n ranging from 1 through $4-6$ than by the simple ${\rm r^{1/4}-}$ law, the latter is still conventionally used for surface brightness decomposition of distant galaxies.}

Despite considerable effort since many years both from the observational side and
from semianalytical and hydrocosmological galaxy formation models, it is not yet
clear, how and when bulges formed. They may either have formed together with the
halo stars and globular clusters prior to the disks or else later from the disks
by internal or external dynamical instabilities such as bar formation and
destruction or an interaction or accretion event.

Among the parameters easiest to access in observational investigations of galaxy
morphologies and their evolution at the limits of present observational
capabilities are the B/D and B/T light ratios. To compare B/D or B/T light
ratios among galaxies at different redshifts in an attempt to study their
evolution it is important to know the wavelength dependence of these light
ratios. From the observational side, a very first systematic comparison of galaxy morphologies obtained from B- and H-band classifications of different spiral types is presented by \citet{esk}. 

In the local Universe, a good correlation of galaxy morphological properties
with their respective spectral and photometric properties is well established.

Spheroidal galaxies and the spheroidal components of spirals, i.e. the bulges,
are well described with star formation laws that declined rapidly from high
values in the distant past to the very low or zero values observed at present.
Evolutionary synthesis models usually use a time evolution of their Star
Formation Rate {\bf (SFR)} of the form ${\rm \Psi(t) \sim exp(-(t/t_{\ast}))}$
with short e-folding times of order ${\rm t_{\ast} \sim 1}$ Gyr (e.g.
\citeauthor{bruz2} \citeyear{bruz2}, \citeauthor{1987A&A...186....1G}
\citeyear{1987A&A...186....1G}, \citeauthor{fritze1} \citeyear{fritze1}, and many others) or
even describe those objects by single burst populations
(\citeauthor{1999ApJ...513..224V} \citeyear{1999ApJ...513..224V},
\citeauthor{2000ApJ...541..126M} \citeyear{2000ApJ...541..126M}).

Star formation in galaxy disks, on the other hand, is a much more steady process
with the SFR density depending on the surface density of HI according to a
Schmidt law ${\rm \Sigma_{SFR} \sim \Sigma_{HI}^n}$ with n close to 1 (cf.
\citeauthor{1998ApJ...498..541K} \citeyear{1998ARA&A..36..189K},
\citeyear{1998ApJ...498..541K}). In evolutionary
synthesis models, therefore, the global SFRs for spiral galaxies are
conventionally ascribed much longer characteristic timescales ranging from ${\rm
  t_{ch} = 2}$ Gyr for Sa to ${\rm t_{ch} \sim t_{Hubble}}$ for pure disk Sd
galaxies.

In the work presented here, we use a very simple and standard evolutionary
synthesis approach to investigate the wavelength dependence of B/T light ratios
in various wavelength bands in their time evolution. We separately model the two
galaxy components, bulge and disk, and then combine them to yield, after 12 Gyr
of evolution, the presently observed average B-band B/T light ratios typical of
different spiral types. We can then study the differences between conventional
B-band B/T ratios and the corresponding light ratios ${\rm (B/T)|_\lambda}$ in
other wavelength bands $\lambda$. We find significant differences in ${\rm
  (B/T)}$ ratios between the B-band where ${\rm (B/T)}$ ratios are conventionally
determined for local galaxies and other wavelength bands.  We show
quantitatively how these differences depend on galaxy type.

Studying our models backward in time, we explore the time and redshift evolution
of ${\rm B/T}$ - ratios in various bands which is important to take into account
when comparing galaxy morphologies at various redshifts.

Finally, we discuss the implications that our results have for the interpretation of the first observational results on the redshift evolution of B/T-ratios.

\section{Models}
\subsection{Evolution of Bulge and Disk Components}

We describe the bulge component by a SFR ${\rm \Psi (t) \sim e^{-t/t_{\ast}}}$
with ${\rm t_{\ast} = 1}$ Gyr and the disk component with a constant SFR ${\rm
  \Psi (t) = \textsl{const}}$. We use our evolutionary synthesis code (cf. \citeauthor{schulz02} \citeyear{schulz02} for details) to
separately calculate the time evolution of spectra, luminosities in various
bands (U ...K), mass-to-light ratios, etc. for both components and then, in a
second step, we combine the two components in mass ratios as to yield, by
today's age, the observed average B-band B/T ratios of the respective galaxy
types (Table \ref{tab:bt_rat})

\begin{table}[!htbp]
\begin{center}
\begin{tabular}{ll}
\hline
Type & ${\rm (B/T)|_B}$ \\
\hline
E  & $\sim 1$ \\
S0 & 0.57 \\
Sa & 0.41 \\
Sb & 0.24 \\
Sc & 0.094 \\
Sd & 0.022 \\
\hline
\end{tabular}
\end{center}
\caption{Average B-band B/T light ratios of local galaxy types from \citet{simien}. }
\label{tab:bt_rat}
\end{table}

We note that both components are modeled in the conventional way, i.e. with
stellar evolutionary tracks/isochrones for solar metallicity. Observational
evidence points to average subsolar metallicities for both bulge and disk stars
(cf. \citeauthor{1994ApJS...91..749M} \citeyear{1994ApJS...91..749M},
\citeauthor{1996AJ....112..171S} \citeyear{1996AJ....112..171S},
\citeauthor{2000AJ....120..833R} \citeyear{2000AJ....120..833R},
\citeauthor{1996MNRAS.279..447R} \citeyear{1996MNRAS.279..447R}, 
\citeyear{1998A&A...339..791R}). At
about half-solar metallicity, a value typical 
for bulge stars and stars in the solar neighborhood, both components bulge and
disk will become slightly bluer ($\sim 0.05$) in their optical color
evolution.

\subsection{Bulge and Disk Formation Scenarios}

Currently, there are basically two scenarios for the formation of bulges and
disks. The first is the early collapse scenario with bulge Star Formation ({\bf
  SF}) occurring on a short timescale $\lta 1$ Gyr from pre-galactic gas early in
the Universe, possibly related to halo star and globular cluster formation.
Disks form later, perhaps after a time delay of a few Gyr from dissipative
infall of left-over halo gas. Elliptical galaxies and the bulges of spirals obey
the same relations e.g. for surface brightness vs. luminosity
\citep{1985ApJ...292L...9K}, magnesium index vs. central velocity dispersion
(e.g. \citeauthor{1993ApJ...411..153B} \citeyear{1993ApJ...411..153B},
\citeauthor{1996AJ....112.1415J} \citeyear{1996AJ....112.1415J}) , suggesting a
common origin for all spheroidal stellar systems.

Large bulges like those of the Milky Way and M31 seem to host old stellar
populations with narrow age dispersions
(\citeauthor{2001ApSSS.277..437P} \citeyear{2001ApSSS.277..437P},
\citeauthor{1999MNRAS.303..641A}  \citeyear{1999MNRAS.303..641A}) in support of this 
early collapse scenario.

In the second scenario, disks form first and bulges are created secularly from
disk material. The onset of bulge formation may have various causes. Internal
instabilities in the disk as well as external perturbations may create bars that
efficiently funnel stars and gas into the inner parts, providing material for
intense SF (e.g. \citeauthor{1997A&A...326..449M}
\citeyear{1997A&A...326..449M}, \citeauthor{1999A&A...351...43A}
\citeyear{1999A&A...351...43A}) . A
high central mass concentration can, in turn, dissolve the bar and leave a bulge
or, in a further step, increase the mass of a pre-existing bulge
\citep{1996ApJ...462..114N}. These processes may happen at any time (and even
several times) in the evolution of a spiral galaxy. In this scenario, part of
the bulge stars may only be spatially regrouped and have an extended age
distribution like that of the disk stars, while others form in the bulge
formation event. Clearly, accretion events where a larger disk galaxy swallows a
smaller companion can also lead to bulge formation. In this case, again, either
a stellar object (dE, dSph) may be accreted and make up the bulge or a gas-rich
dwarf galaxy that may fuel a central starburst producing the stars in the bulge
of the later remnant.  While a mere regrouping of pre-existing disk stars would
give the bulge stars a broad age distribution, the formation of a significant
population of new stars in the bulge-forming event will occur on short
timescales ($10^8 - 10^9$ yr) and leave a bulge with a stellar population
younger than that of the disk but with a narrow age distribution.

Some recent observational results seem to favor the view that while large bulges
were formed according to the early collapse scenario, i.e. on a short timescale
in the early Universe with the disks around them starting SF later, small
bulges may form later than the disks during their secular
evolution and by internal instabilities or external accretion
(\citeauthor{1994AJ....107..135B} \citeyear{1994AJ....107..135B},
\citeauthor{2001A&A...366...68P} \citeyear{2001A&A...366...68P}, see
also \citeauthor{1999Ap&SS.267..145W} \citeyear{1999Ap&SS.267..145W} for a recent
review).

For the very abundant numbers of S0 galaxies in local galaxy clusters, yet a different formation scenario is discussed. Cluster S0 galaxies are found to 
be most plausibly transformed into what they are from the spiral-rich field galaxy population falling into the cluster potential. Spectroscopic analyses by \citet{jo2000}, our own comparison of their photometric properties with evolutionary synthesis models including SF truncation with or without prior starbursts \citep{bicker2002} and dynamical modelling (\citeauthor{mo1998} \citeyear{mo1998}, \citeauthor{qu2000} \citeyear{qu2000}) together draft a scenario where rapid interactions with the cluster potential and environment largely destroy the stellar disks of infalling spirals and the dense and hot intracluster medium sweeps away their HI supply, ultimately leaving remnants with S0 morphologies and SF completely truncated. This, however, is not the subject of the present paper where we want to explore the effects of bulge versus disk SF along the Hubble sequence. 

\subsection{Combining Disks and Bulges}

To cope with the uncertainty about the evolutionary sequence of bulge and disk
formation, we investigate 3 different scenarios. In the first one, both
components are assumed to start SF at the same time, i.e. $\sim 12$ Gyr ago. In
the second one, only the bulge starts forming stars that early, at a redshift
${\rm z \sim 3 - 5}$ in standard cosmologies (we use ${\rm H_o = 65,~ \Omega_o =
  0.1,~ \Lambda_o = 0}$) and SF in the disk is assumed to only start with some
delay about 7 Gyr ago, i.e. at ${\rm z \sim 1}$. In the third scenario, the disk
is assumed to start SF first and bulge formation is assumed to take place in a
secondary SF event about 7 Gyr ago, like a central starburst as e.g. following a
disk instability, an external perturbation of the disk, or an accretion event.

We just chose one redshift ${\rm z \sim 1}$ for the onset of delayed bulge or disk SF and postpone a detailed study of this parameter to a future paper that directly compares models with observational B/T-decompositions.

Note that our models do not include any dynamical effects, they are simple
1-zone models without spatial resolution. They only go beyond classical global
galaxy evolutionary synthesis models by the inclusion of two phases of SF for the disk and
the bulge which are superposed at appropriate times.

\section{Spectral Evolution}

\begin{figure*}[!htb]
\hbox{
\scalebox{0.9}{\includegraphics[width=6.5cm]{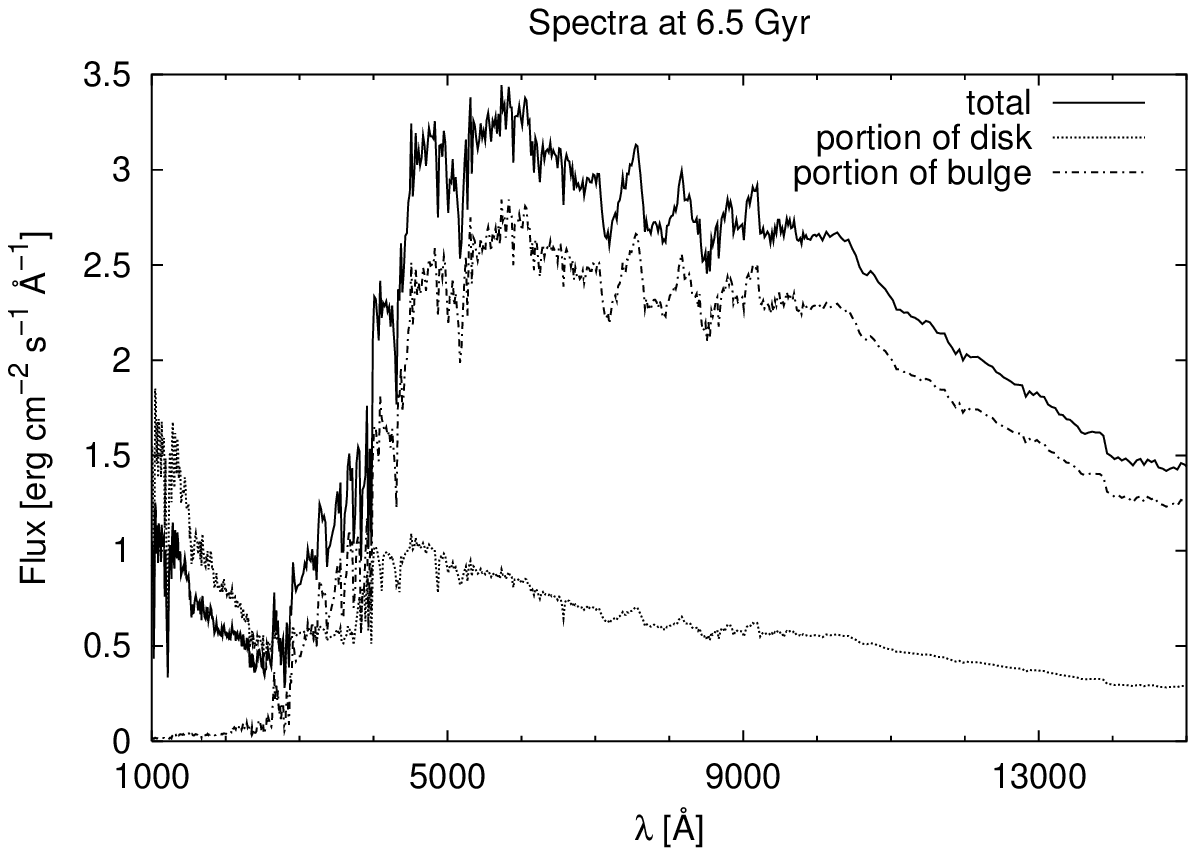}}
\scalebox{0.9}{\includegraphics[width=6.5cm]{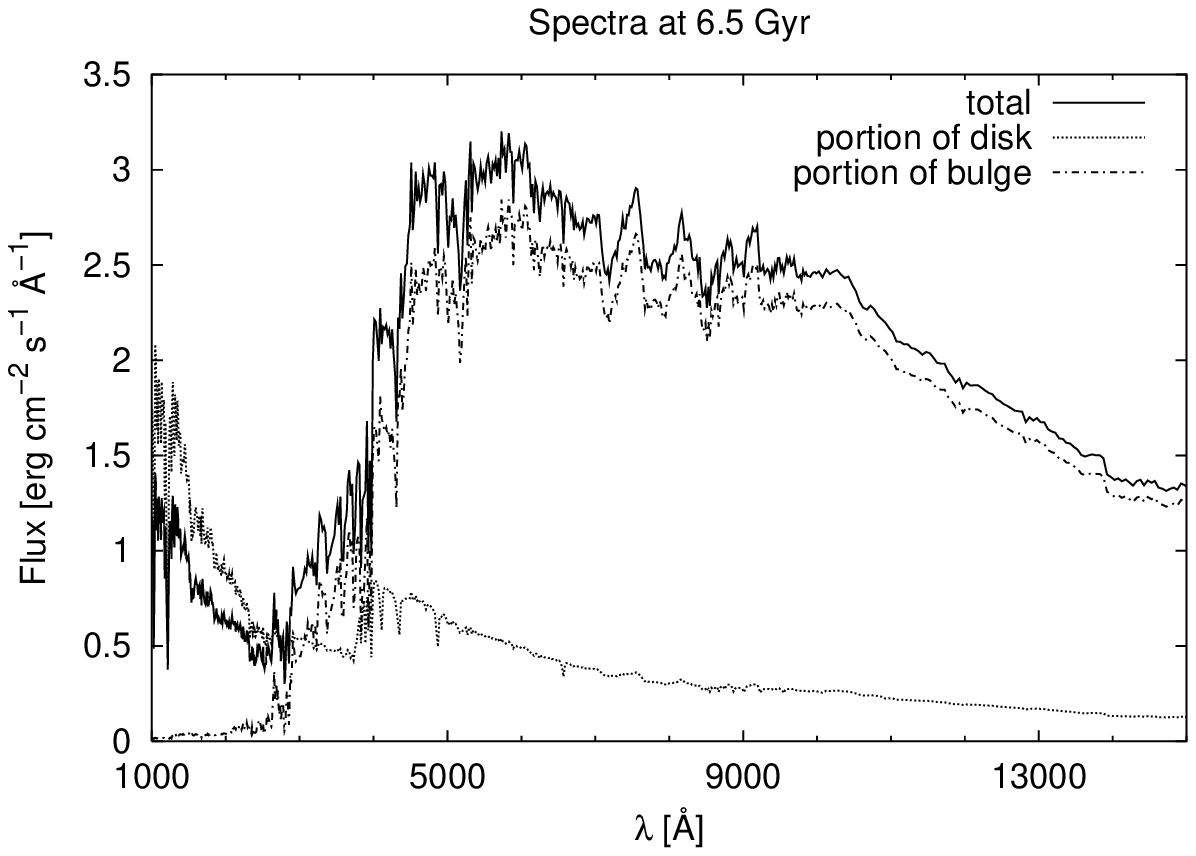}}
\scalebox{0.9}{\includegraphics[width=6.5cm]{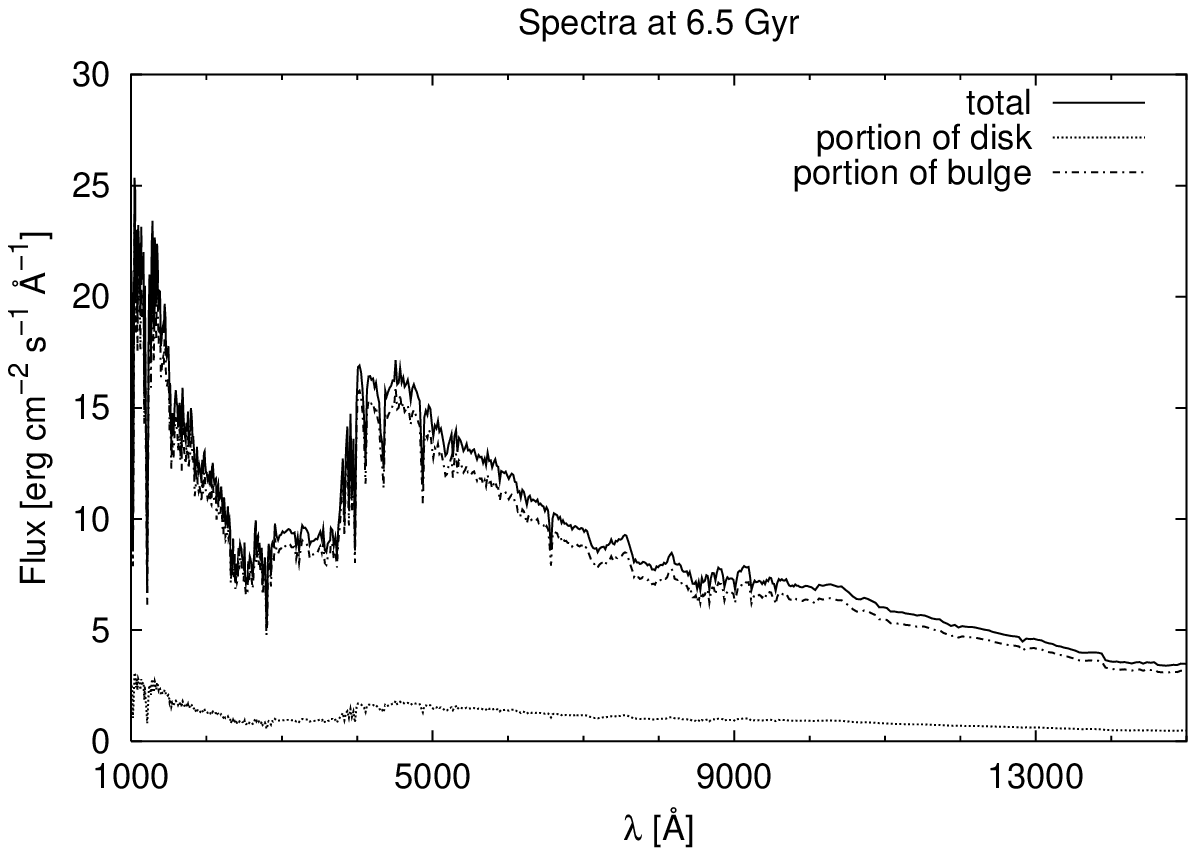}}
}
\hbox{
\scalebox{0.9}{\includegraphics[width=6.5cm]{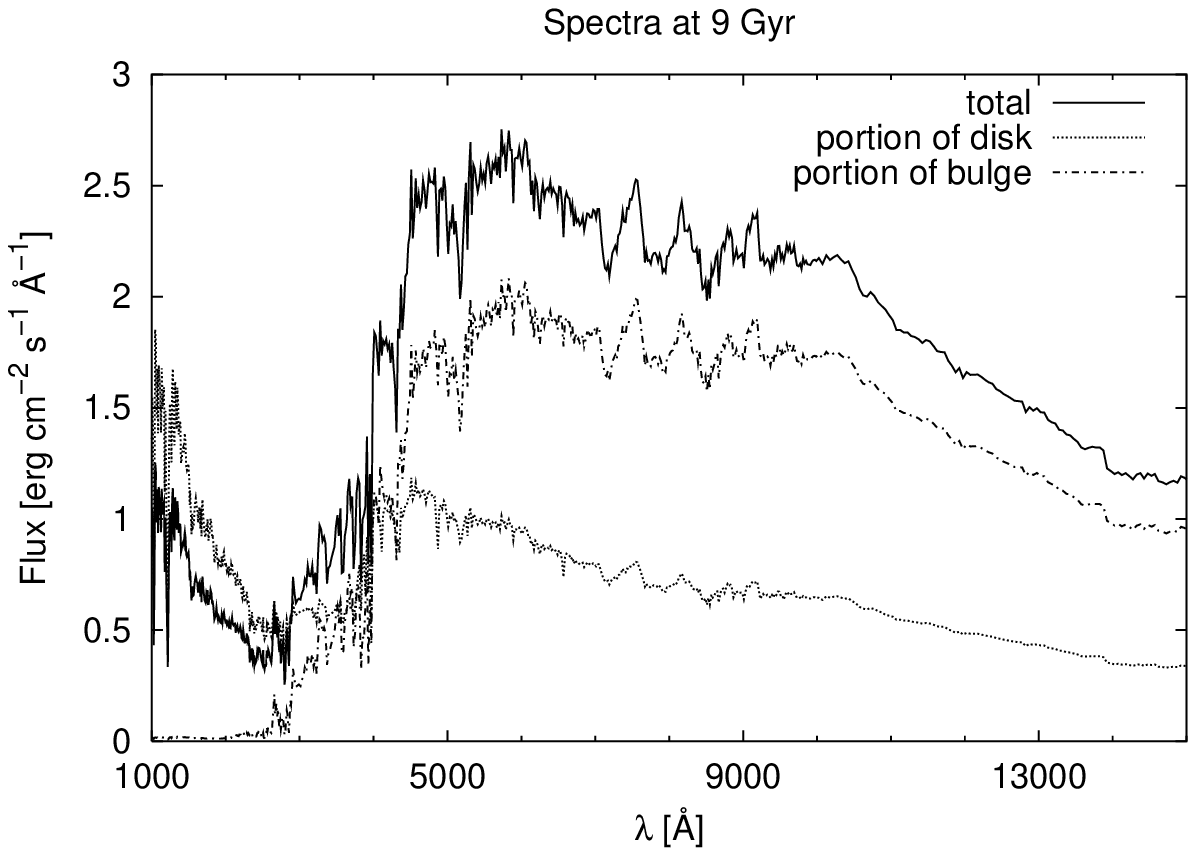}}
\scalebox{0.9}{\includegraphics[width=6.5cm]{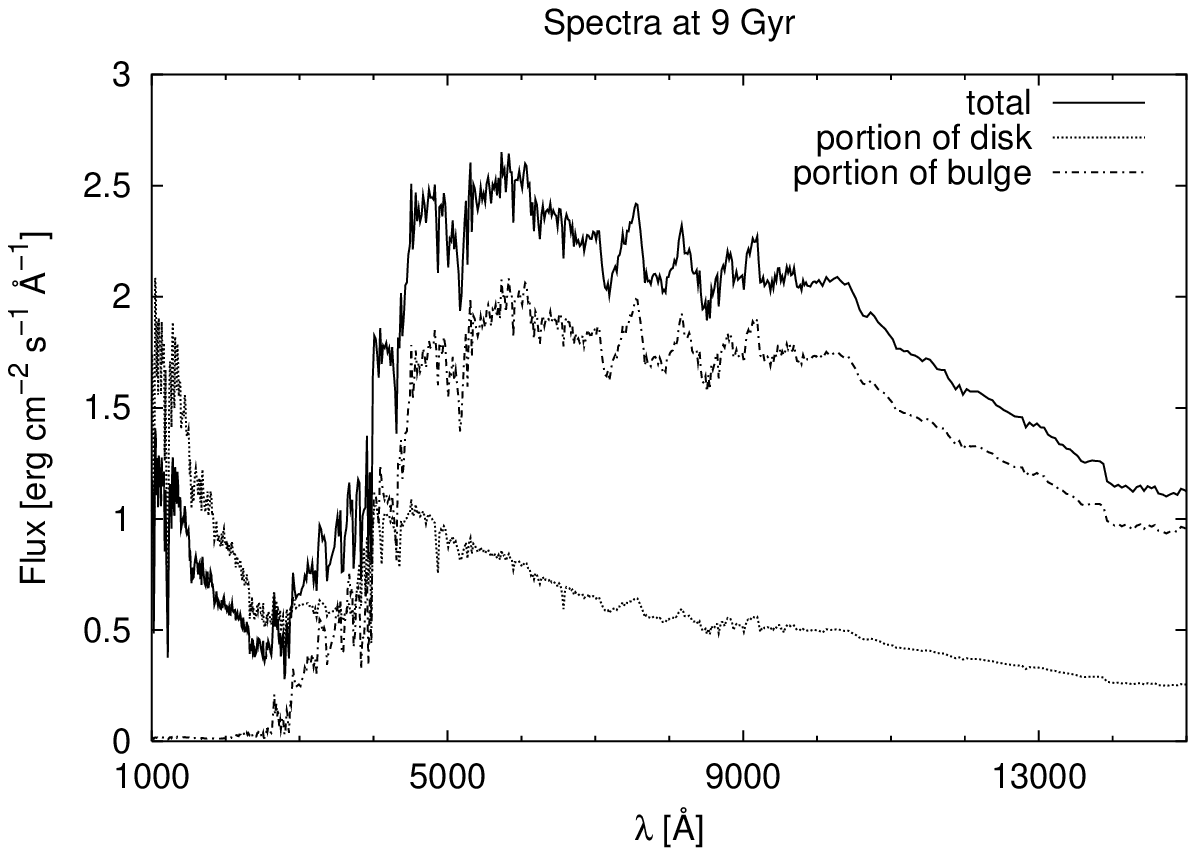}}
\scalebox{0.9}{\includegraphics[width=6.5cm]{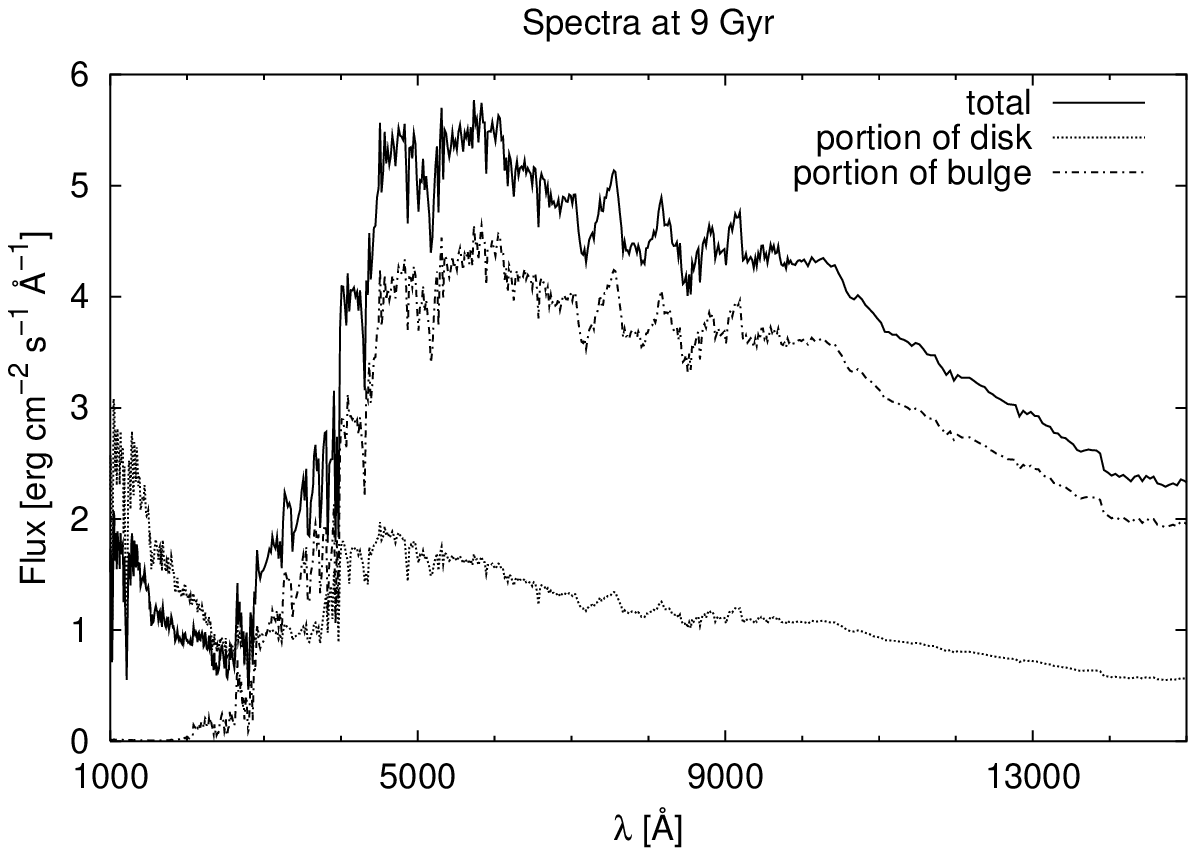}}
}
\hbox{
\scalebox{0.9}{\includegraphics[width=6.5cm]{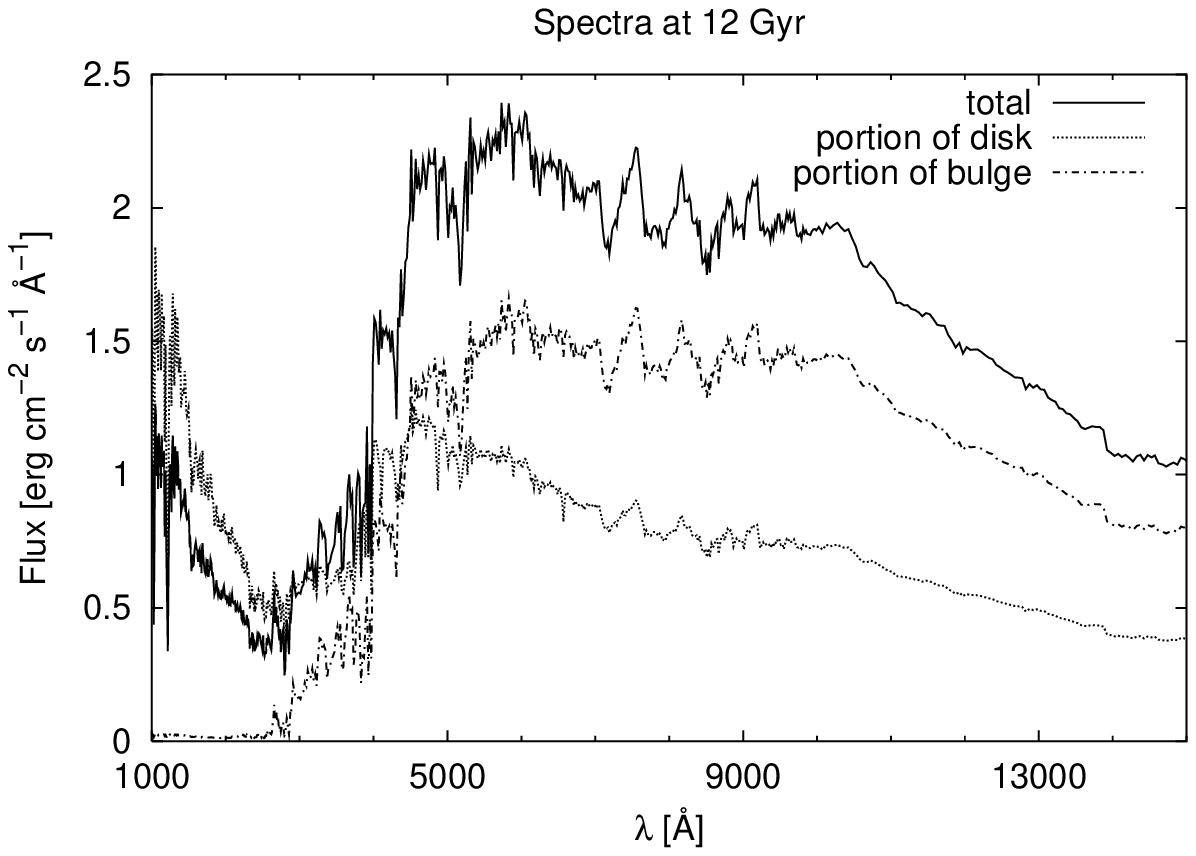}}
\scalebox{0.9}{\includegraphics[width=6.5cm]{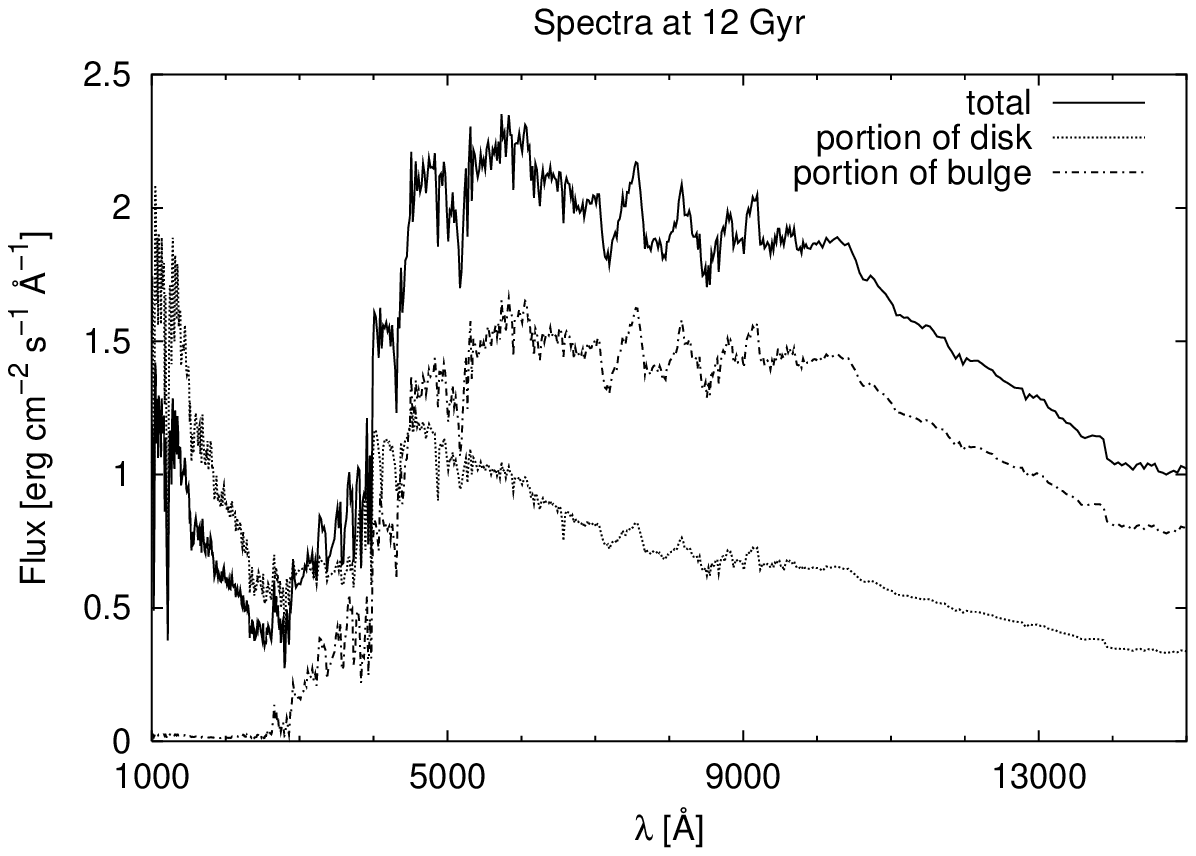}}
\scalebox{0.9}{\includegraphics[width=6.5cm]{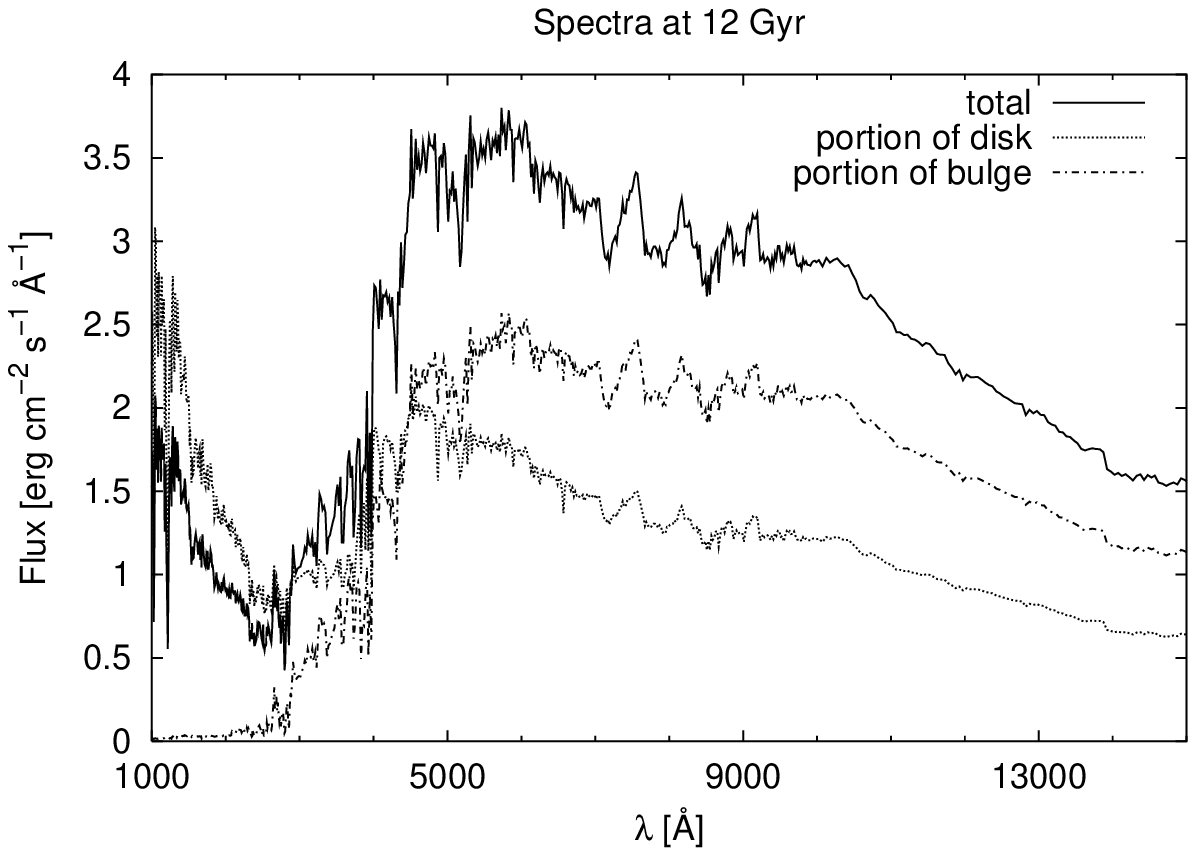}}
}
\caption{Spectra of a galaxy with present-day ${\rm (B/T)|_B = 0.6}$ at ages of
6.5, 9, and 12 Gyr, respectively. The first column shows a model with equal age 
bulge and disk components, the second column the model with an old bulge and a 
younger disk, and the third column the model with old disk and a younger bulge.}
\label{fig:oo_s}
\end{figure*}

In Fig. \ref{fig:oo_s} we present the spectral evolution of a galaxy with
present-day ${\rm (B/T)|_B = 0.6}$, classified as S0, in the three different
scenarios (Col. 1 to 3) at three different ages ranging from 6.5 Gyr (${\rm
  1^{st}}$ row) through 9 and 12 Gyr (${\rm 2^{nd}}$ and ${\rm 3^{rd}}$ row,
respectively). Each panel presents the integrated spectrum of the galaxy from
the UV through NIR (solid line) and the decomposition into bulge (dashed line)
and disk components (dotted line).

The absolute luminosities of the galaxies at 12 Gyr are the same in all three
scenarios. The earlier the bulge forms, the higher its early SF rate has to be
to lead to a fixed ${\rm (B/T)|_B = 0.6}$ at 12 Gyr, the later disk SF starts,
the higher its constant rate has to be.

The normalization to ${\rm B/T = 0.6}$ in the B-band at 12 Gyr is clearly seen
in terms of nearly comparable fluxes of the bulge and disk components at ${\rm
  \lambda \sim 4400}$ \AA.

For the equal age bulge and disk scenario presented in the ${\rm 1^{st}}$ Col. we see that at early times $\sim 6.5$ Gyr, the bulge with its strong early SF
strongly dominates the spectrum at optical and NIR wavelengths. Bulge SF at this
age, however, is essentially finished already and disk SF is largely responsible
for the UV flux. As both components age, the UV contribution of the disk
component forming stars at a constant rate is essentially unchanged since the
number of high mass stars has reached an equilibrium. Towards longer
wavelengths, the flux from the disk slightly increases with age as low mass
stars accumulate. The flux from the bulge component decreases as bulge stars go
on dying out from higher to lower masses.

In the scenario with the old bulge and a disk starting SF only around 5 Gyr
(${\rm 2^{nd}}$ Col.), the disk contribution at optical and NIR wavelengths at
an age of 6.5 Gyr is still fairly small with the bulge dominating very strongly,
because of his many small mass stars. Only in the UV again, where the old bulge
has not much flux any more at this age, the disk component, although still very
small in mass, clearly dominates because of their newly born massive stars. To end
up with the same ${\rm (B/T)|_B = 0.6}$ as in the first scenario, the disk now
has to have a higher constant SFR and is seen to catch up in optical and NIR
fluxes at ages 9 and 12 Gyr like in the scenario shown in Col. 1.

In the ${\rm 3^{rd}}$ Col. showing the scenario of an old disk and delayed
bulge SF, the situation is very different. At 6.5 Gyr we see a very young bulge
shortly after its violent SF phase that largely dominates the flux from UV
through NIR. Note the high overall luminosity of a galaxy with such a young
bulge component. Around 5000 {\AA} e.g. the total flux in this scenario is
higher by a factor 3 as compared to both other scenarios. At 9 Gyr, the total
flux at 5000 {\AA} is only higher than in the other scenarios by a factor $\sim
1.3$, bulge SF has ceased and the UV has become dominated by the disk component
which also gains relative importance at longer wavelengths, a trend that
continues through the age of 12 Gyr.

\section{Photometric Evolution}

\begin{figure*}[!htb]
\hbox{
\scalebox{0.9}{\includegraphics[width=6.5cm]{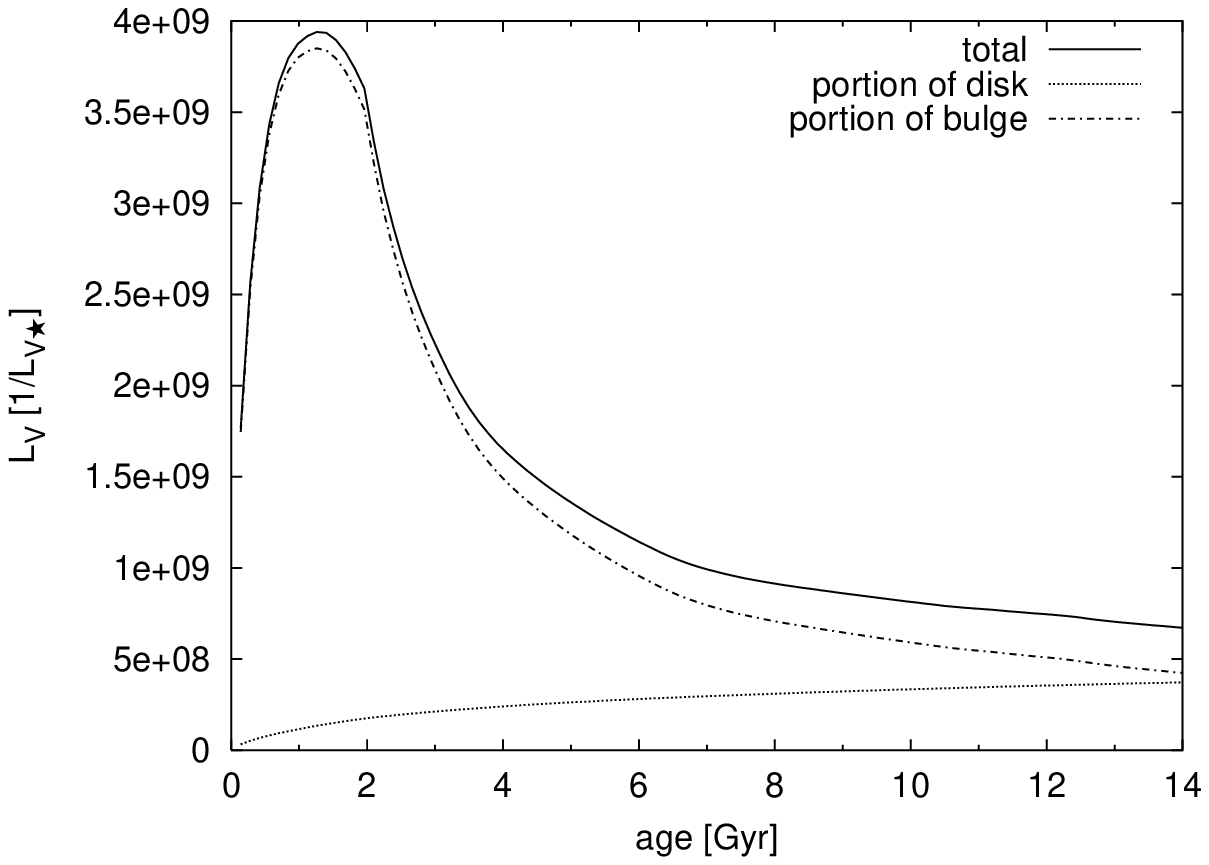}}
\scalebox{0.9}{\includegraphics[width=6.5cm]{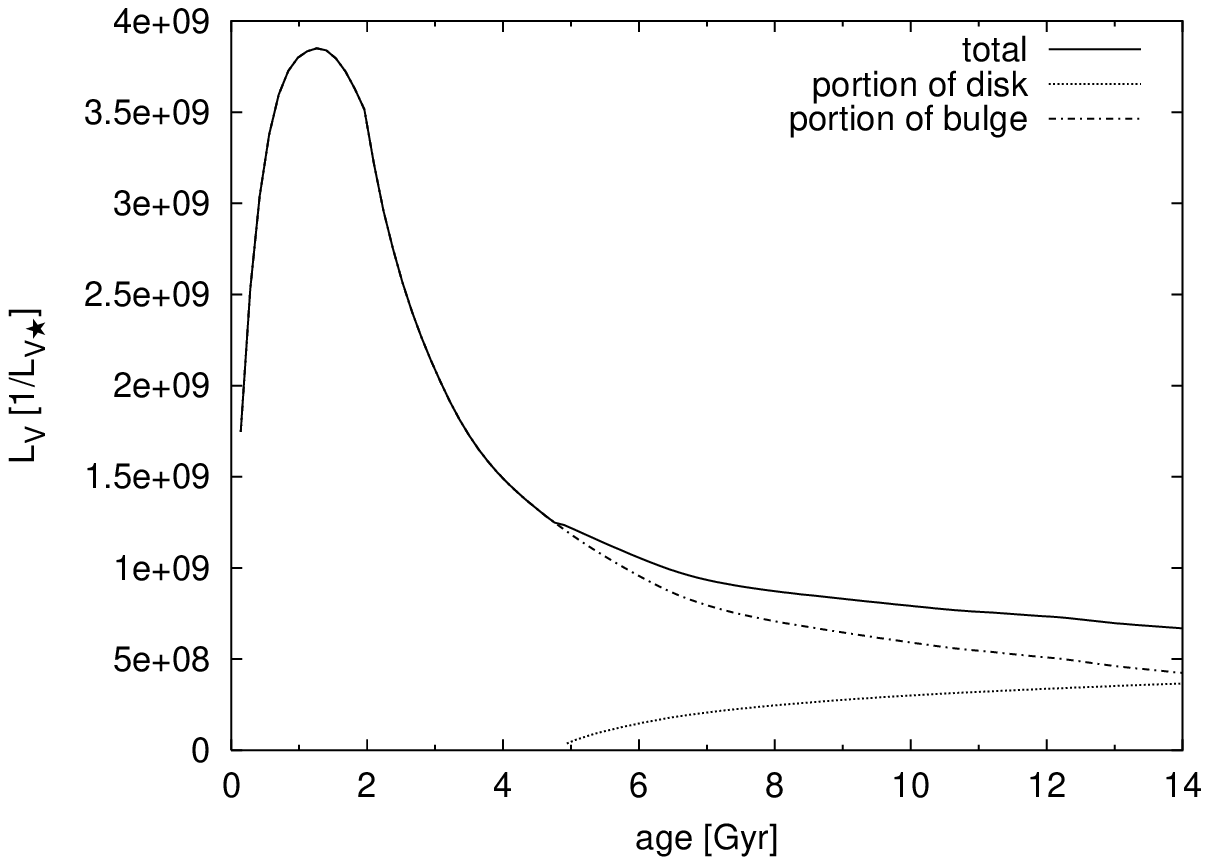}}
\scalebox{0.9}{\includegraphics[width=6.5cm]{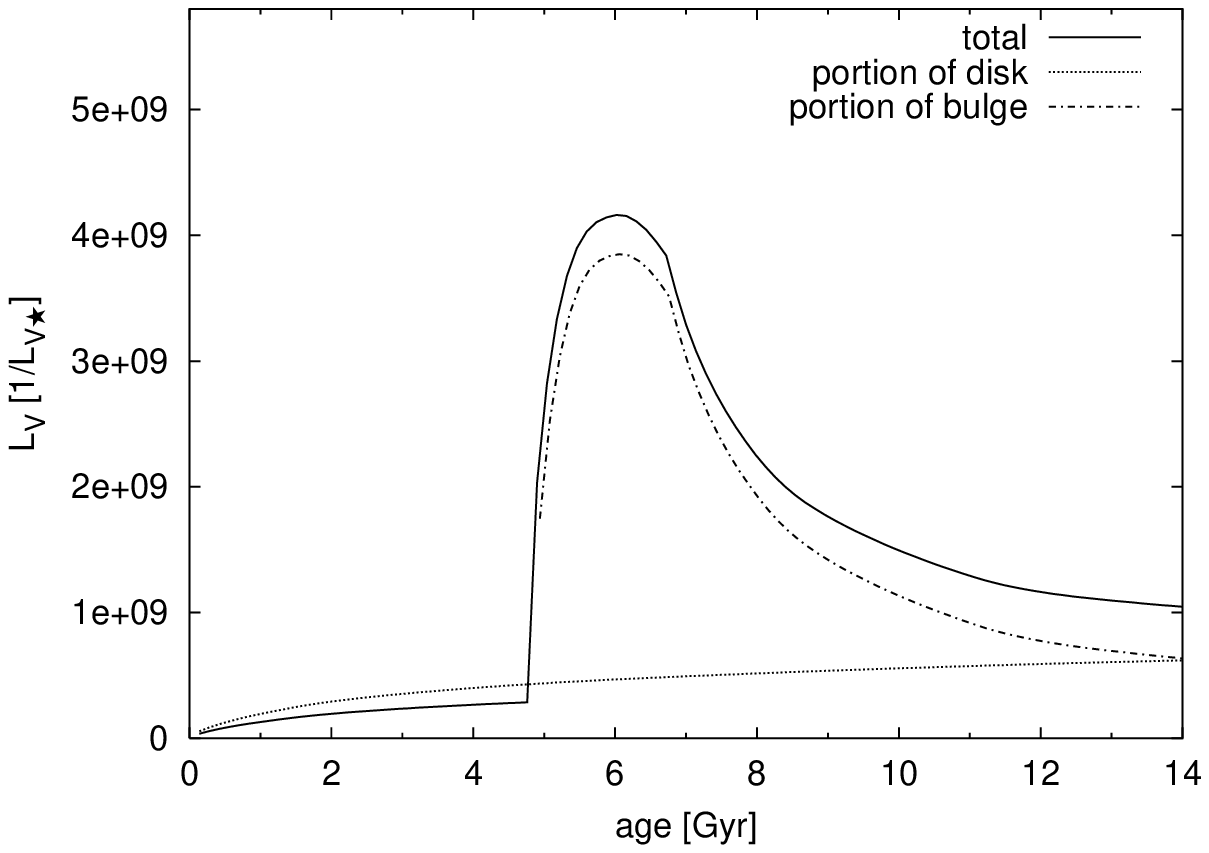}}
}
\hbox{
\scalebox{0.9}{\includegraphics[width=6.5cm]{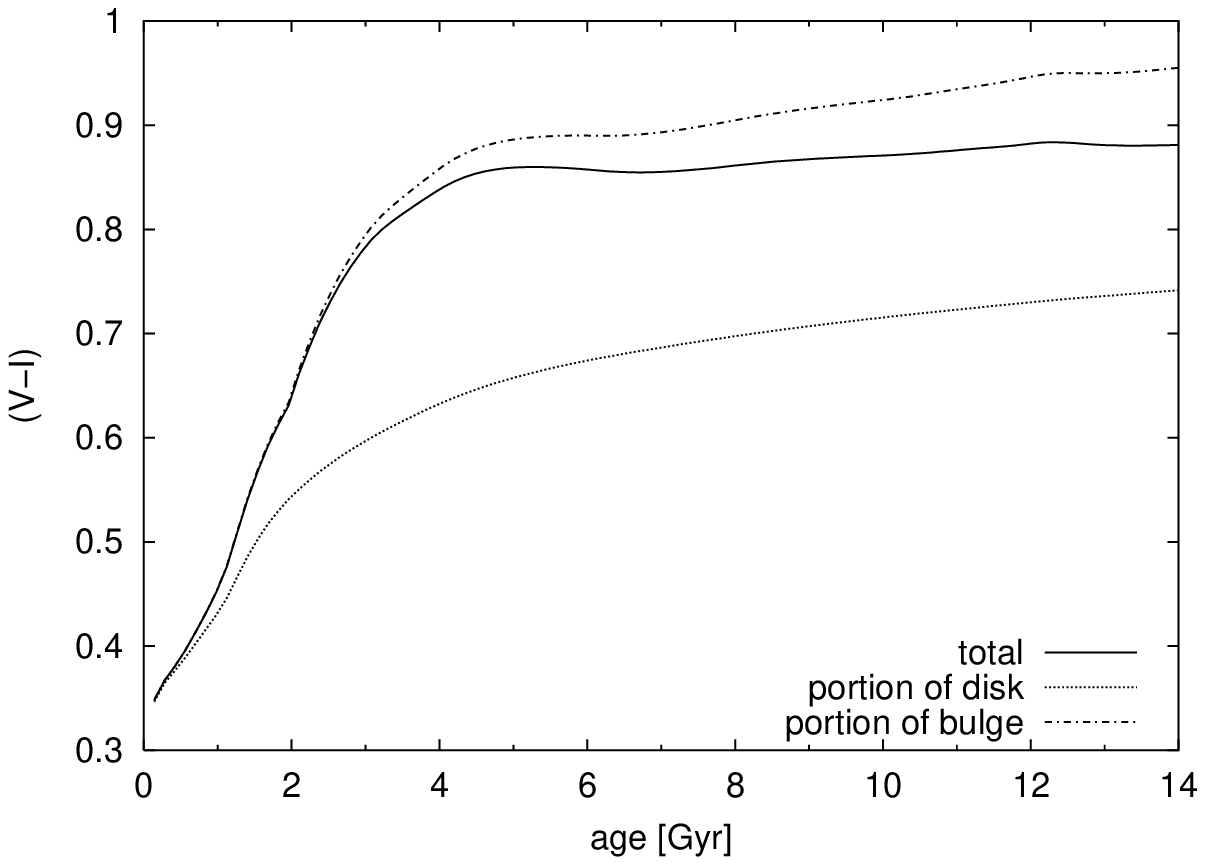}}
\scalebox{0.9}{\includegraphics[width=6.5cm]{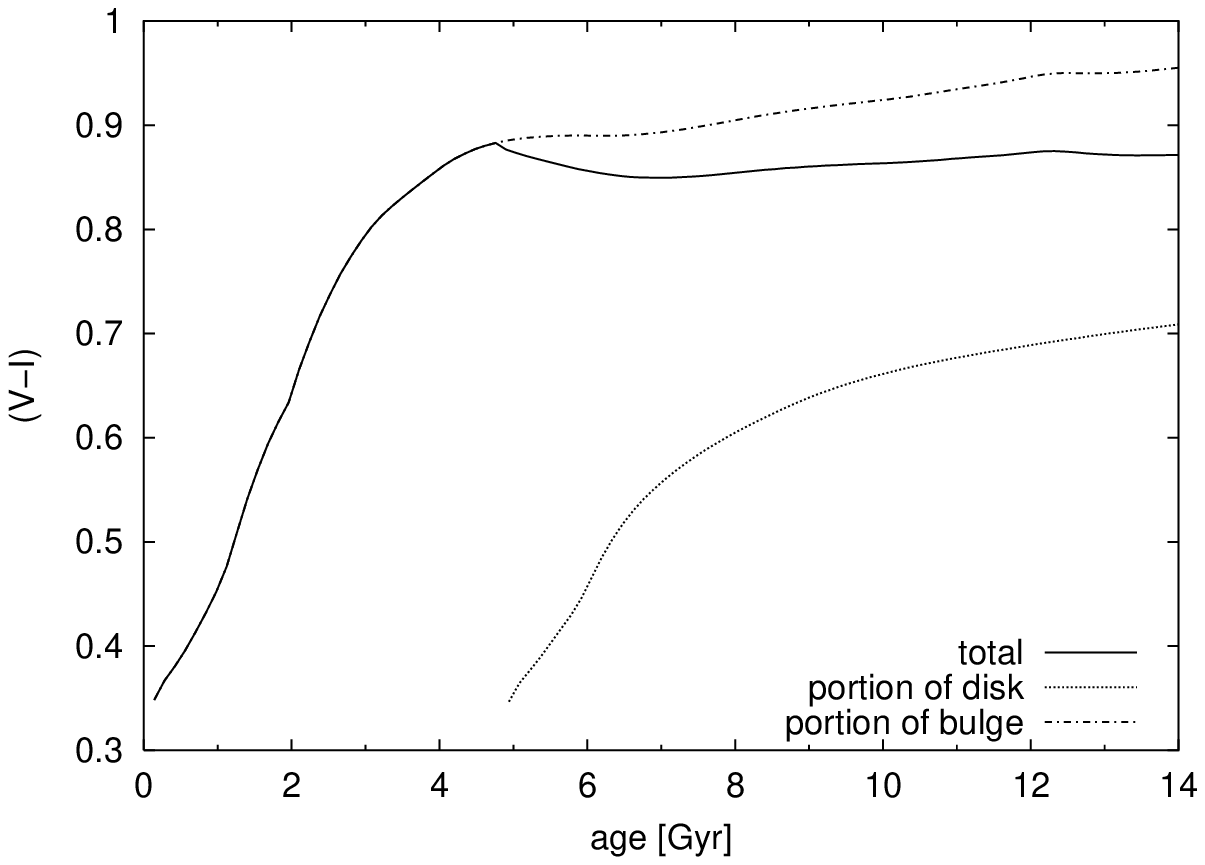}}
\scalebox{0.9}{\includegraphics[width=6.5cm]{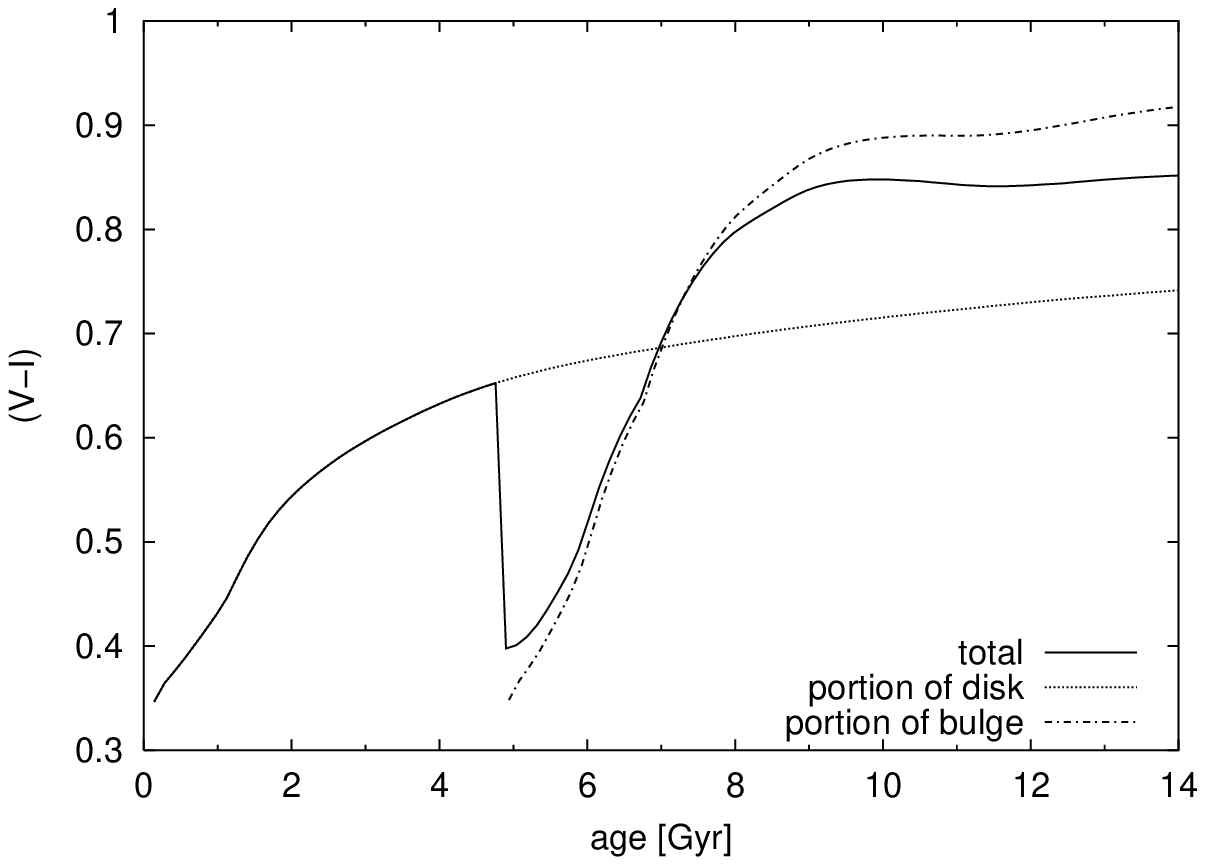}}
}
\caption{V-band luminosity (upper panels) and ${\rm (V-I)}$ color evolution (lower panels) for model galaxies with present-day ${\rm (B/T)|_B = 0.6}$. The left column again shows the model with equal ages of
bulge and disk, the middle column shows the model with an old bulge and a younger disk, and the right column the model with an old disk and a younger bulge.}
\label{fig:oo-v}
\end{figure*}

In this Section we present the photometric evolution of galaxies with
present-day ${\rm (B/T)|_B = 0.6}$ in the three different scenarios. In
Fig. \ref{fig:oo-v} we show as an example the V-band luminosity and (V-I) color
evolution for the total system (solid lines) and the decomposition into bulge
and disk components (dotted and dashed lines, respectively). Evolution is shown
from the onset of SF through the present time, corresponding to $\sim 12$ Gyr,
and beyond. Hence, we zoom into the V-band wavelength region of
Fig. \ref{fig:oo_s} to study it in some more detail.  Of course, this can be
done for all other wavelengths from UV through NIR, but we will present only one
example here.

We see in all the top panels of Fig. \ref{fig:oo-v} the strong V-band luminosity
evolution associated with the burst-like bulge SF (${\rm \Psi \sim
  e^{-t/1 \, Gyr}}$) producing a stellar population that after strong fading over
12 or 7 Gyr, for the old and younger bulge scenarios, respectively, still contributes 60 \% of the total light in B today. The
burst-like bulge SF leads to a strong increase in V-luminosity shortly after the
onset of bulge SF and a strong decline starting $\gta 1$ Gyr later. The
continuous disk SF mode, on the other hand, goes along with a weak but steady
rise in disk V-luminosity from the onset of disk SF to the present.

For the total V-luminosity evolution it hardly makes any difference if disk SF
starts as early as bulge SF or 5 Gyr later. In both the old bulge/younger disk
and equal age cases, the V-luminosity is much higher (about a factor 4) in early
evolutionary stages ($\lta 3$ Gyr) than today. Only in the old disk/younger
bulge case the system starts out at significantly lower V-band luminosity than
today until the onset of the bulge SF sharply raises its V-luminosity by a factor $>
10$. About $\lta 2$ Gyr later, it starts fading again.

Looking at the ${\rm (V-I)}$ color evolution (lower panels) we see, in the case
of equal age bulge and disk components, a significant reddening by 0.5 mag
during the first 4 Gyr and a constant ${\rm (V-I) \sim 0.88}$ for all times
thereafter. The bulge component further reddens at ${\rm t > 4}$ Gyr, but this
is compensated by the increasing importance of the bluer disk.

For the old bulge and younger disk case, the total color evolution, of course,
only shows the strong reddening of the bulge component to ${\rm (V-I) \sim 0.9}$
until, at ${\rm t \gta 5}$ Gyr, disk SF sets in and acts to bring the ${\rm
(V-I)}$ back to 0.85 within $\sim 2$ Gyr from whereon it remains $\sim$
constant.

In the old disk and younger bulge case, as expected, the system reddens much
less and only reaches ${\rm (V-I) \sim 0.65}$ after 5 Gyr of evolution. The
onset of the strong burst-like bulge SF then induces a strong and sudden blueing
to ${\rm (V-I) \sim 0.4}$ which then turns into a steady reddening to reach
${\rm (V-I) \sim 0.85}$ about 3 Gyr later.

Differences in today's total ${\rm (V-I)}$ color for a galaxy with present-day ${\rm
(B/T)|_B = 0.6}$ are small between the different scenarios. At 12 Gyr,
${\rm (V-I) = 0.87}$ in the case of equal age components, and 0.85 and 0.83 for
older bulge and older disk scenarios, respectively.

\section{Wavelength Dependence of Present Day B/T - Ratios for Various Galaxy Types}
\subsection{Model Results}
\begin{figure}[!ht]
\includegraphics[width=\columnwidth]{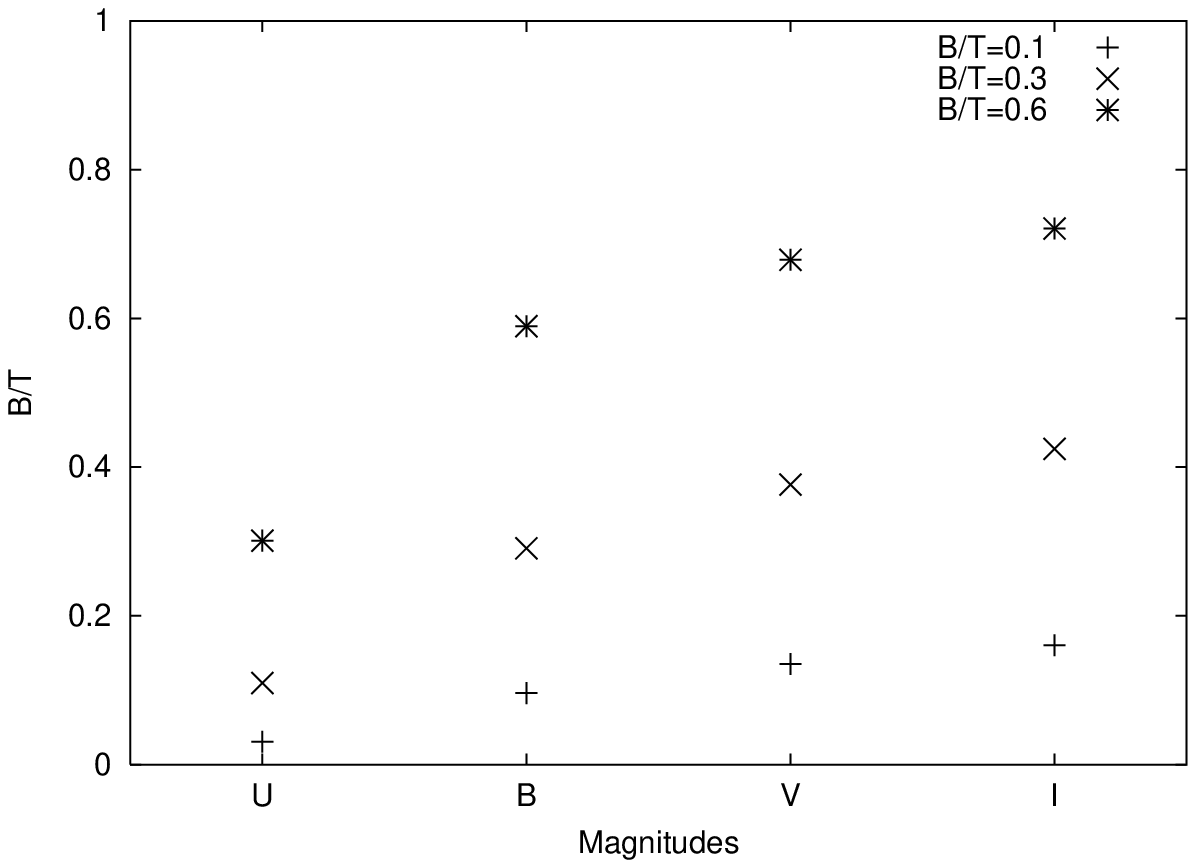}
\includegraphics[width=\columnwidth]{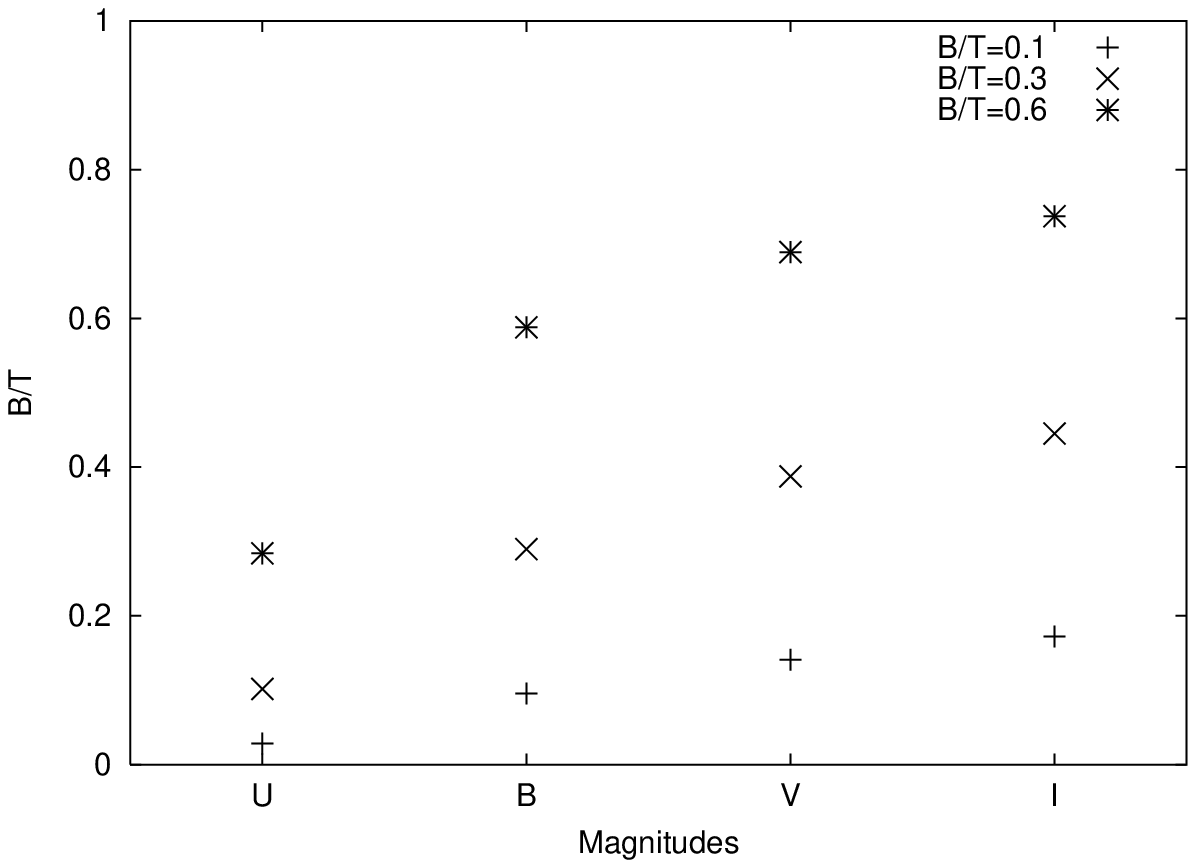}
\includegraphics[width=\columnwidth]{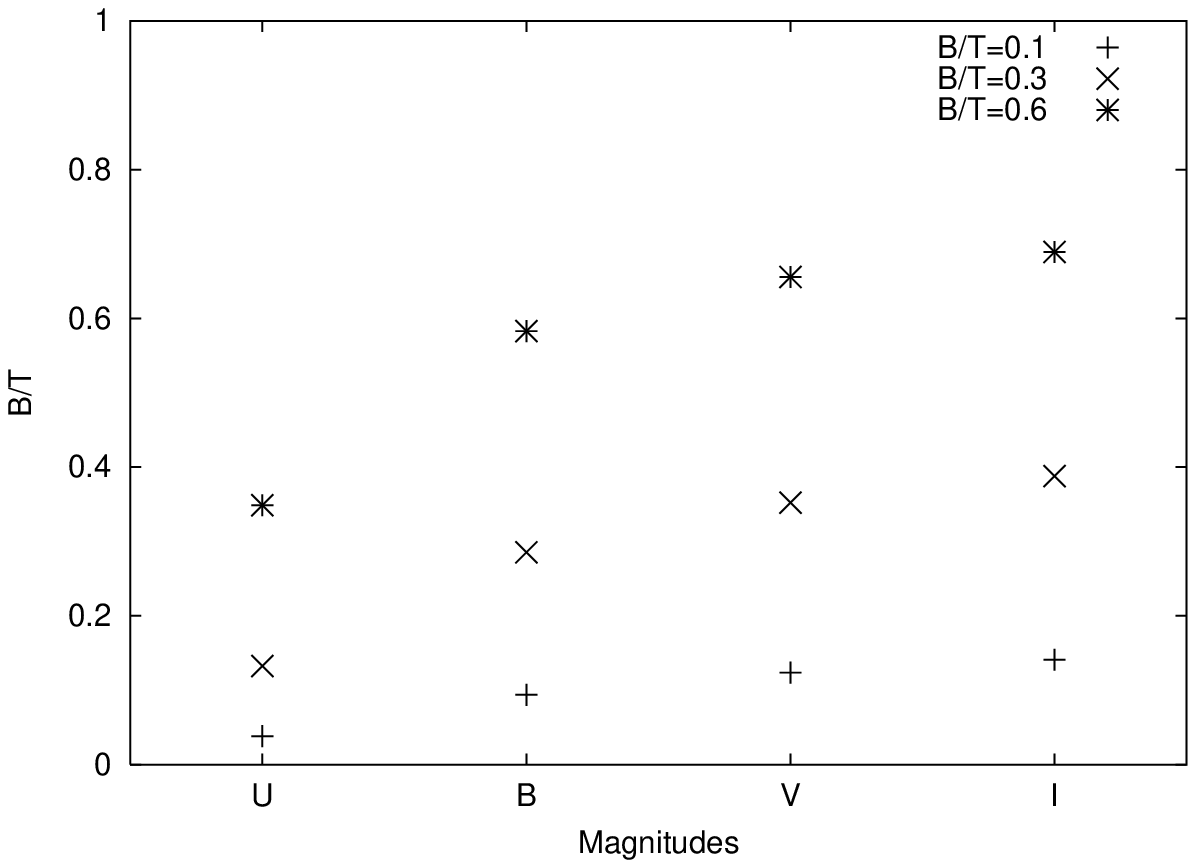}
\caption{Present-day ${\rm B/T}-$ ratios in different filters for model 
galaxies with ${\rm (B/T)|_B = 0.1,~0.3,~and~0.6}$, respectively. Top 
panel: equal age components, middle panel: old bulge and younger disk, 
bottom panel: old disk and younger bulge.}
\label{fig:bt_mag}
\end{figure}

In this Section, we investigate the wavelength dependence of B/T ratios for
galaxies of various types, i.e. with various present-day B-band ${\rm (B/T)|_B}$
ratios of 0.1 ($\leftrightarrow$ Sc), 0.3 ($\leftrightarrow$ Sab), and 0.6
($\leftrightarrow$ S0). In Fig. \ref{fig:bt_mag}, we present the wavelength
dependence of ${\rm (B/T)|_{\lambda}}$ values for these three galaxy types in
all three scenarios. All galaxies here are depicted at their present age of 12
Gyr. Note that we use HST filter bands for UBVI throughout. 

For all galaxies, there is a clear trend for B/T ratios to increase with
increasing wavelength of observation. There is little difference between the
scenarios with equal age bulge and disk components and with old bulges and
younger disks.  E.g., an Sc-type galaxy with B-band ${\rm (B/T)|_B = 0.1}$ shows
${\rm (B/T)|_U = 0.03}$, ${\rm (B/T)|_V = 0.13}$, and ${\rm (B/T)|_I = 0.17}$. Its
B/T ratio hence increases by a factor $\sim 6$ from U through I in both scenarios
with an old bulge.

\begin{table}[!htbp]
\begin{center}
\begin{tabular}{lllll}
\hline
Type & ${\rm (B/T)|_U}$  & ${\rm (B/T)|_B}$  & ${\rm (B/T)|_V}$  & ${\rm (B/T)|_I}$ \\
\hline
\multicolumn{5}{c}{equal age bulge and disk }\\
E  & 0.72 & 0.9 & 0.93 & 0.94 \\
S0 & 0.30 & 0.6 & 0.68 & 0.72 \\
Sa & 0.11 & 0.3 & 0.38 & 0.42 \\
Sc & 0.03 & 0.1 & 0.14 & 0.16 \\
\multicolumn{5}{c}{old disk}\\
E & 0.76 & 0.9 & 0.92 & 0.93\\
S0 & 0.35 & 0.6 & 0.66 & 0.69\\
Sa & 0.13 & 0.3 & 0.35 & 0.39\\
Sc & 0.04 & 0.1 & 0.12 & 0.14\\
\multicolumn{5}{c}{old bulge}\\
E & 0.70 & 0.9 & 0.93 & 0.94\\
S0 & 0.28 & 0.6 & 0.69 & 0.74\\
Sa & 0.10 & 0.3 & 0.39 & 0.45\\
Sc & 0.03 & 0.1 & 0.14 & 0.17\\
\hline
\end{tabular}
\end{center}
\caption{Wavelength dependence of B/T light ratios for galaxies of different
type, i.e. with different present-day B-band B/T ratios. We only show the case of 
equal age bulge and disk components.}
\label{tab:bt_wav}
\end{table}


Table \ref{tab:bt_wav} shows that the trend for B/T to increase with increasing
wavelength is stronger for the small bulge galaxies than for those with strong
bulge components.

For the old disk and younger bulge scenario, however, the trend of increasing
B/T with increasing wavelength is weaker. From ${\rm (B/T)|_U = 0.04}$ through
${\rm (B/T)|_I = 0.15}$ it only increases by a factor $\sim 4$.

All of this is due to the different SF histories and timescales of the disk and
bulge components. Disk SF being essentially constant in time from its onset to
the present epoch soon produces a stellar population in which the fraction of high
mass stars dominating the wavelength region shortward of V keeps constant and
the low mass stars dominating longward of V slowly accumulate. Bulge SF being
strongly peaked around the time of bulge formation implies that after a very
blue high luminosity period bulge stars become collectively old and tend to
dominate the longer wavelength regions. Their effect, of course, is the
stronger, the higher the final required B/T ratio of the galaxy.

{\bf We conclude} that as a result of the different SF timescales for bulge and
disk components the B/T light ratios of galaxies are predicted to be
significantly dependent on the wavelength at which the decomposition is
performed. Slightly dependent on the galaxy formation scenario, they
systematically increase from U through I with the strongest effect, a factor of
4 -- 6 difference from U to I, seen for the weak bulge galaxy types.

In principle, bulge -- disk decomposition of local galaxies at
different wavelengths may give information about the relative ages of the bulge
and disk stars and, hence, help constrain bulge and disk formation scenarios.

\subsection{First Comparison with Observations}
\citet{esk} recently presented a first systematic comparison of galaxy classifications in B and H. They decompose their sample of 250 bright spirals of various types from S0 through Sm from the Ohio State University Bright Spiral Galaxy Survey in B- and H-bands independently. They find that, while for extremely bulge-strong (S0) as well as pure disk systems (Sd...Sm) the classifications from B and H do not significantly diverge, spirals from Sa through Scd appear one T-type earlier in H than in B, on average, albeit with large scatter. We see from Fig. 3 and Table 2 that in I our Sab and Sc models models have B/T-ratios very similar to the B-band B/T of Sa and Sbc models, respectively, not much dependent on the scenario adopted and well consistent with Eskridge \etal's results. As in their observations, our S0 model with ${\rm (B/T)|_B=0.6}$ does not change much when observed in I: ${\rm (B/T)|_I\sim 0.7}$. The large observational scatter precludes us from definitely discriminating between the three different scenarios for bulge and disk formation, although the observations seem to better agree with old bulge models than with the one where the disk is older than the bulge. Their results may, as well, indicate that there is in fact a range of ages for real disks (and bulges) rather than just the two extremes that we study. A detailed quantitative analysis of these data with our models, however, will be the subject of a forthcoming paper. 

\section{Redshift Evolution of B/T - Light Ratios} 

Comparing galaxy morphologies at different redshifts is a difficult task. Two
effects have to be considered, the band-shift effect, related to the
cosmological correction, resulting from the redshifting and dimming of the
spectrum, and the evolutionary effect due to changes in the SF rate.

As to the first effect, it is immediately clear from a comparison of images of
local galaxies at different wavelengths, e.g. in the UV, optical, and NIR, that
galaxy morphologies appear significantly different at different
wavelengths. E.g., star-forming regions or knots stand out in the UV. To describe
quantitatively how galaxy morphologies change due to this band shift effect when
going towards higher redshift, however, is very difficult.

To quantify evolutionary effects on the morphological appearance of an
individual galaxy as a function of redshift would require to know its individual
SF history and, in particular, the SF histories of its components bulge and
disk. While the SF histories of individual galaxies may to some extent be
different from galaxy to galaxy, in particular in their short term fluctuations,
there seem to be reasonable average global SF histories, smoothed out over short
term fluctuations, that are characteristic of galaxies of various spectral
types. This is a very basic fact on which all evolutionary synthesis models
rely. And, in the local Universe, at least, a good correlation is observed
between spectral and morphological galaxy types. How far this correlation holds
back into the past is an open question and depends on the formation scenario for
the bulge and disk components of galaxies.

Careful analysis of observations of galaxy morphologies out to significant
redshifts, that are becoming possible now with HST and large ground-based
telescopes, are a promising way to investigate these questions. \citet{mar}
present B/T ratios from an ${\rm r^{1/4}-}$ bulge and exponential disk
decomposition in the I-band (F814W) of more than 500 galaxies brighter than
${\rm I = 26}$ mag in the HDF North over the redshift range ${\rm 0.1 \lta z
  \lta 4.5}$. They find the number of bulge dominated galaxies to decrease with
increasing redshift. The apparent lack of pure disk galaxies at ${\rm z \gta 3}$
in their data may be due to a selection effect driving the low surface
brightness disk galaxies below the detection/classification limit at those
redshifts. The lack of high surface brightness strong bulge components at redshifts ${\rm z \gta 1}$, however,
cannot be explained by a selection bias. A detailed analysis of the relative
contributions of bulge and disk components, derived from I-band images over this
large redshift interval requires consideration of both band-shift and
evolutionary effects.

As simplified as they are, our bulge and disk models offer a first tool for
comparison with results from bulge -- disk decompositions of distant
galaxies. In this Section, we present a theoretical analysis of the redshift
evolution of B/T ratios as obtained from our models for the three different
galaxy formation scenarios we consider. As for the wavelength dependence of
present-day B/T ratios discussed above, the redshift evolution of ${\rm
(B/T)|_{\lambda}}$ ratios, as well, is a result of the different SF timescales
that apply for spheroidal and disk galaxy components, respectively. The results
we present fully account for both the cosmological and evolutionary corrections
as explained above with the cosmological model ${\rm H_0 = 65,~\Omega_0 =
0.1,~\Lambda_0 = 0}$. We stress that UBVI note observer-frame (not 
rest-frame) passbands throughout.

\subsection{Bulge and disk components of equal age}
\label{sect:oo}
Let us look first at the scenario where both bulge and disk components were
forming early in the Universe at ${\rm z \sim 5}$.

\begin{figure}[!ht]
\includegraphics[width=\columnwidth]{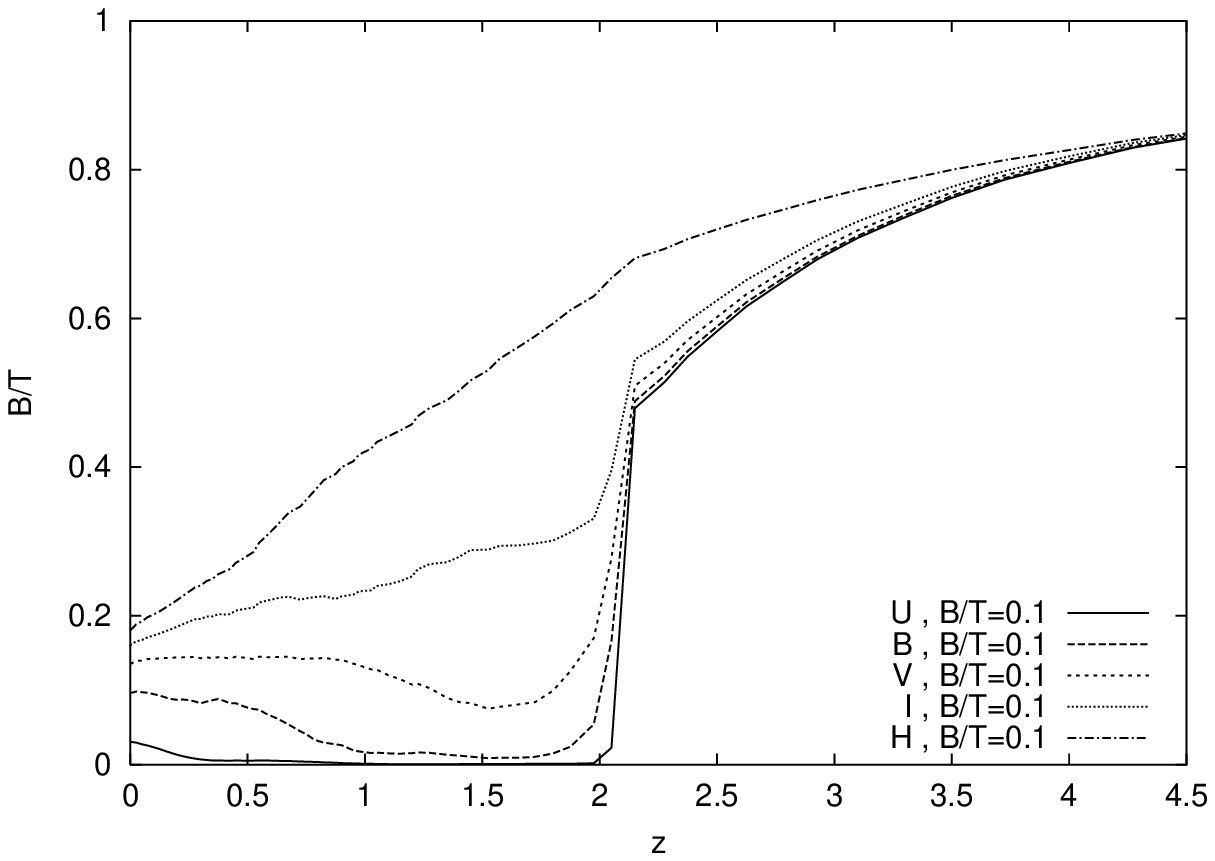}
\includegraphics[width=\columnwidth]{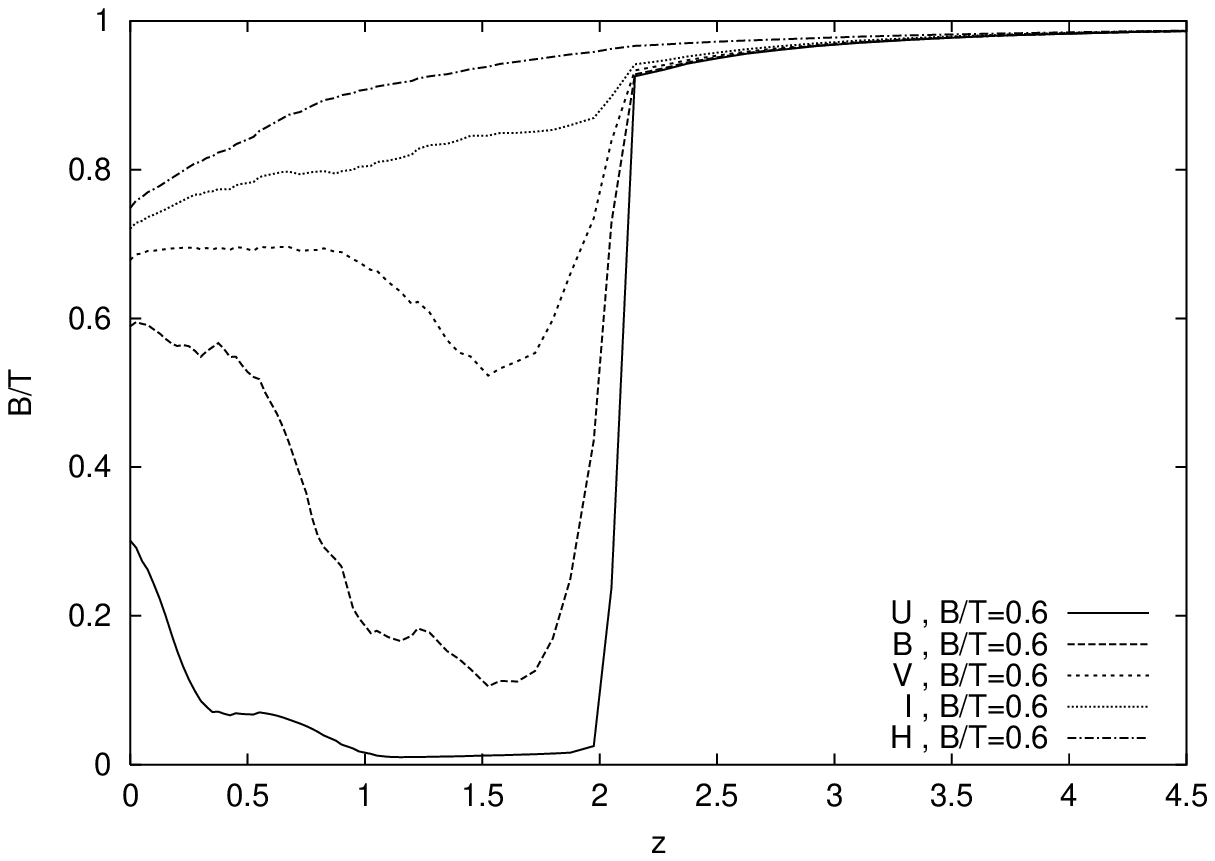}
\caption{Redshift evolution of B/T - ratios in various bands for two different 
galaxy types with present-day ${\rm (B/T)|_B = 0.1}$ (top) and ${\rm (B/T)|_B = 0.6}$ (bottom), respectively. 
Equal age bulge and disk components.}
\label{fig:bt_m_oo}
\end{figure}

Here the situation is dramatic for the progenitors of both types of galaxies,
those ending up today with small and large bulge contributions. In Fig.
\ref{fig:bt_m_oo}, we present the redshift evolution of B/T ratios at ${\rm
  \lambda =}$ U, B, V, I, H out to ${\rm z > 4}$ for two types of galaxies ending
up today at ${\rm z = 0}$ with low and high (B/T)${\rm|_B = 0.1 ~
  (\leftrightarrow~Sc)}$ and 0.6 ${\rm (\leftrightarrow~ S0)}$, respectively. 
  
At present, no H-band decompositions into bulge and disk light contributions of distant galaxies have been published. As HST NICMOS observations, however, will soon provide a wealth of spatially resolved data (Balcells 2002, {\sl priv. comm.}), we include our model H-band predictions in Figs 4 - 6 and discuss them in Sect. 6.4.

For the {\bf late-type galaxies} we first recover at ${\rm z = 0}$ the difference of a
factor $\lta 2$ between its B- and I-band B/T ratios. This difference increases
to a factor $> 20$ out to ${\rm z \sim 2}$ due to both a decrease in ${\rm
(B/T)|_B}$ and an increase in ${\rm (B/T)|_I}$. At ${\rm z \gta 2}$,
corresponding to ages $\lta 2$ Gyr in our cosmology, the active bulge SF phase
acts to rapidly raise the B/T ratios in all bands to a uniform value of 0.5 from
whereon they synchronously continue to increase towards ${\rm (B/T) \sim 0.85 ~
at ~ z \sim 4.5}$.

The progenitors of present-day low B/T galaxies, e.g. with ${\rm (B/T)|_B (z=0) =
0.1 ~(\leftrightarrow ~ Sc)}$ show systematically higher ${\rm (B/T)|_I}$ ratios
at higher redshift, ranging from (B/T)${\rm|_I \sim 0.2 - 0.3}$ at ${\rm z \lta 2}$ to
(B/T)${\rm|_I \sim 0.55 - 0.85}$ at z ${\rm \sim 2.2 - 4.5}$. Hence, including evolutionary and cosmological effects, we find the I-band B/T-ratios of present-day late-type galaxies to rise strongly with redshift.

For {\bf bulge-dominated galaxies} with (B/T)${\rm|_B = 0.6}$ at ${\rm z = 0}$, the
situation is even more extreme in this scenario. Out to ${\rm z \gta 1.5}$, their
B-band B/T ratios decrease strongly down to (B/T)${\rm|_B \sim 0.1}$ at ${\rm z
\sim 1.5}$ as a result of the Balmer discontinuity being redshifted towards longer passbands. Their I-band B/T ratios are not affected and slowly increase.

For strongly bulge-dominated galaxies uniformly classified in I over the huge
redshift range from ${\rm 0 \leq z < 3}$, in this scenario, no
significant biases are expected except that bulge contributions are slightly
overestimated at ${\rm z > 1}$.

In this scenario, galaxies observed at ${\rm z \lta 2}$ should show vastly
different B/T ratios in different bands, varying between ${\rm (B/T)|_U \sim
  0~and~(B/T)|_I \sim 0.3}$ for Sc progenitors and between ${\rm (B/T)|_U \sim
  0.2~and~(B/T)|_I \gta 0.8}$ for S0 progenitors.

Galaxies at ${\rm z > 2}$ are predicted to show the same B/T ratios in all bands
U -- I and only high values ${\rm B/T > 0.5}$ are expected to be observed at
${\rm z > 2}$ since even low mass bulges in their early evolutionary phases ($
\lta 2$ Gyr) strongly dominate the light at all wavelengths. Galaxies observed
at high redshift $\gta 2$ with B/T ratios of 0.5 and greater may well be the
progenitors of local Sc galaxies (${\rm (B/T)|_B = 0.1}$). They simply are
observed in or shortly after their active bulge SF phase.

\subsection{Old bulge and younger disk}

\begin{figure}[!ht]
\includegraphics[width=\columnwidth]{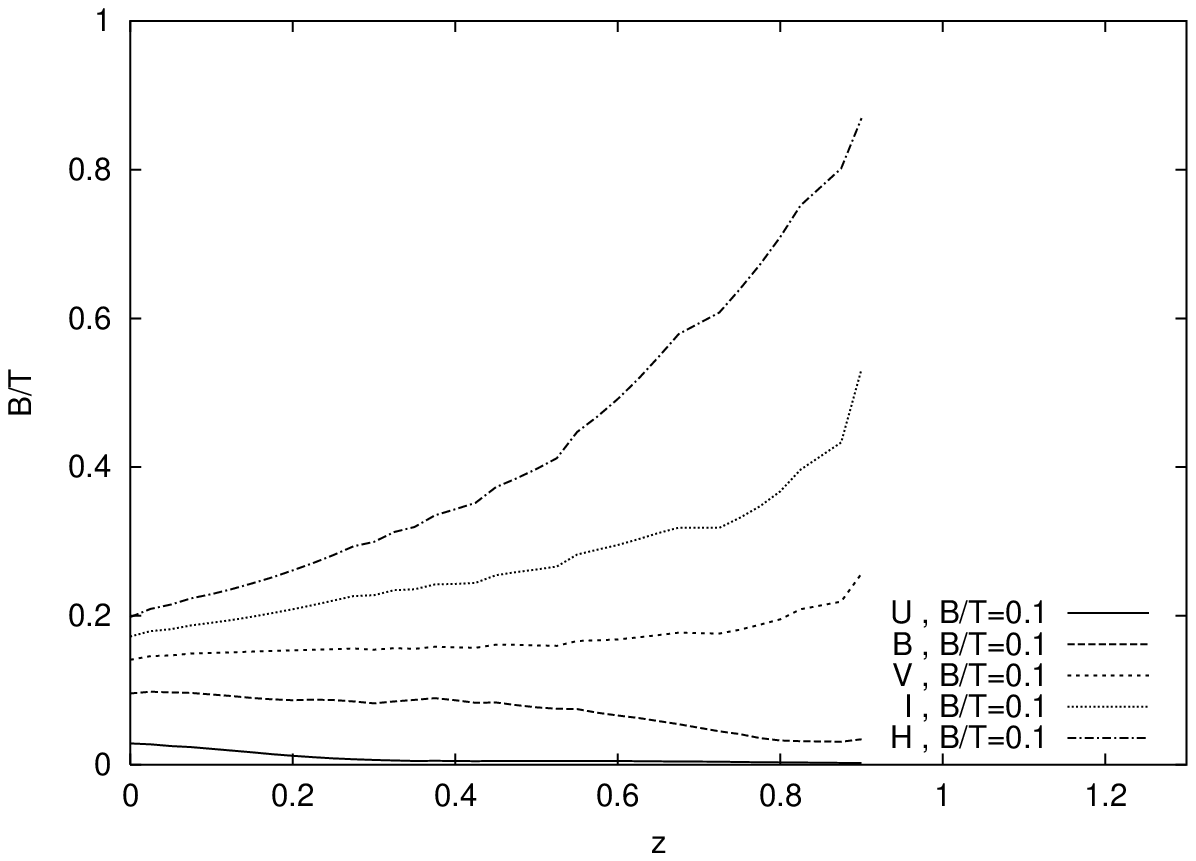}
\includegraphics[width=\columnwidth]{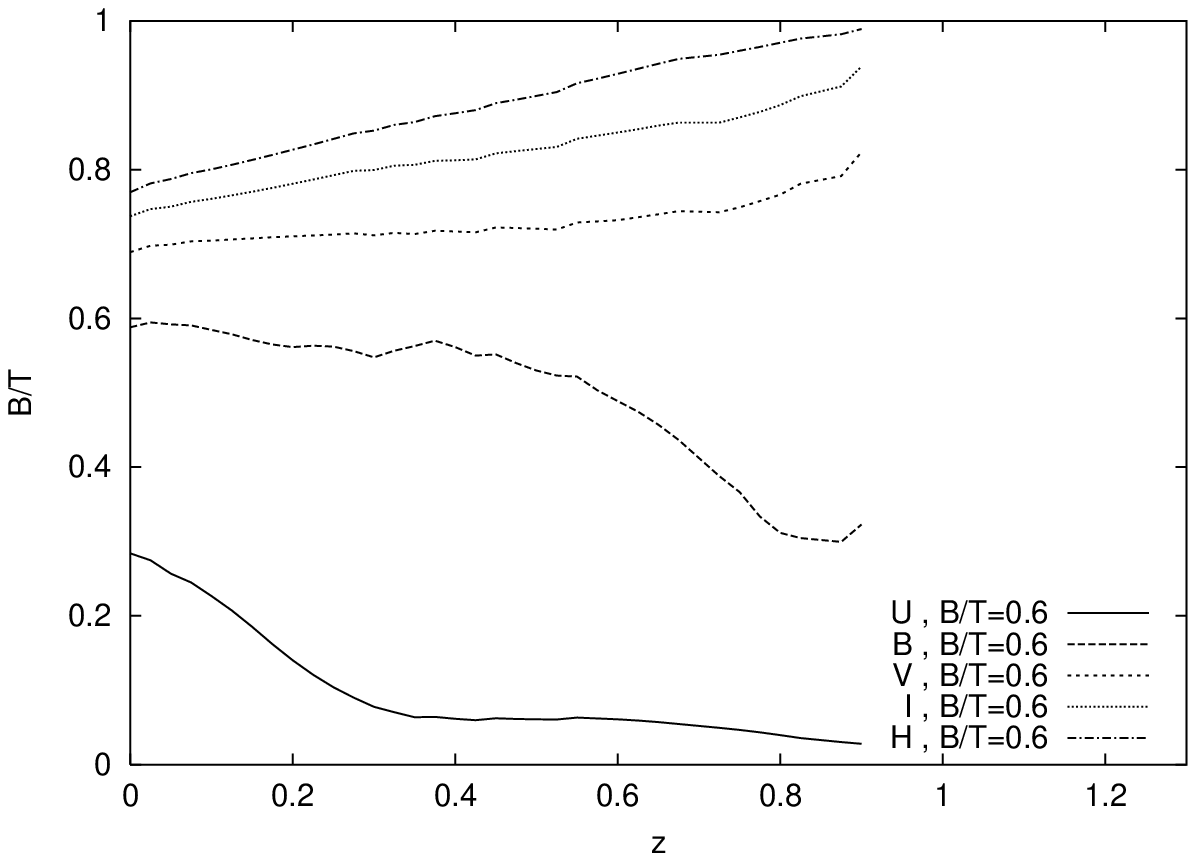}
\caption{Redshift evolution of B/T - ratios in various bands for two different galaxy 
types with present-day ${\rm (B/T)|_B = 0.1~and~0.6}$, respectively. Old bulge and 
younger disk.}
\label{fig:bt_m_ob}
\end{figure}

In Fig. \ref{fig:bt_m_ob}, we present the redshift evolution of B/T ratios in
various bands U, B, V, I, H for two different galaxy types with present-day B-band
${\rm (B/T)|_B (z=0) = 0.1}$ (${\rm \leftrightarrow~Sc}$) and 0.6 (${\rm \leftrightarrow~S0}$), respectively, for the scenario with
bulge formation in the early Universe and disk formation delayed until ${\rm z
  \sim 1}$.

In this scenario, the evolution of low B/T systems is very weak and smooth until
${\rm z \sim 0.9}$. At ${\rm z \gta 0.9}$, the disk did not form stars,
hence only pure bulges are expected. With increasing redshift, the difference
between B- and I-band B/T ratios increases steadily from a factor $\sim 2$ at
${\rm z = 0}$ to a factor $> 10$ at ${\rm z \gta 0.8}$. For high B/T systems, we again observe with increasing redshift a decrease of the U- and B-band B/T-ratios like in the previous scenario and a weak increase of the V- and I-band B/T-values. 

With I-band classification over the redshift range ${\rm 0 \leq z \leq 0.8}$, a 
galaxy ending up with ${\rm (B/T)|_B = 0.1}$ at ${\rm z = 0 ~ (\leftrightarrow ~ 
Sc)}$ would be observed to have ${\rm (B/T)|_I \sim 0.2}$ at ${\rm 0 \leq z \lta 
0.2 ~ (\leftrightarrow ~ Sbc)}$ and ${\rm (B/T)|_I \sim 0.4 }$ around 
${\rm z \sim 0.8 ~ (\leftrightarrow ~ Sa)}$. 

Strong differences
are seen between B/T ratios in different bands for late galaxy
types. Within a given band, however, the redshift evolution is moderate to ${\rm z \sim 0.8}$.

A galaxy ending up with a strong bulge ${\rm (B/T)|_B = 0.6}$ at ${\rm z = 0~(\leftrightarrow S0)}$
would be observed in the I-band to even have a somewhat stronger one at ${\rm 0.3 \lta z \lta 0.9}$.

Hence, in this scenario the progenitors of local weak bulge systems, when
classified in I at redshifts ${\rm 0.2 \lta z \lta 0.9}$ look like having slightly stronger bulge components and this effect tremendously increases, of course, as
one approaches the redshift where disk SF slowly started.

\subsection{Old disk and younger bulge}

\begin{figure}[!ht]
\includegraphics[width=\columnwidth]{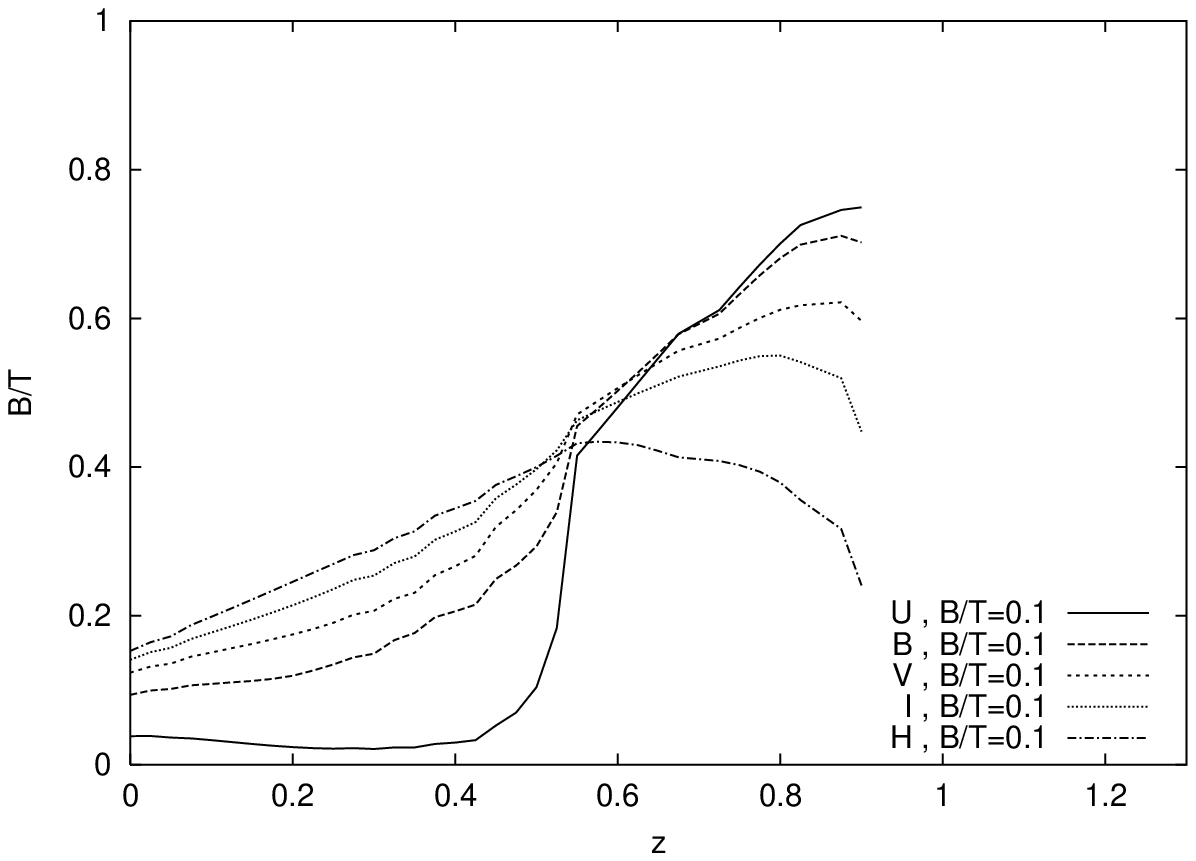}
\includegraphics[width=\columnwidth]{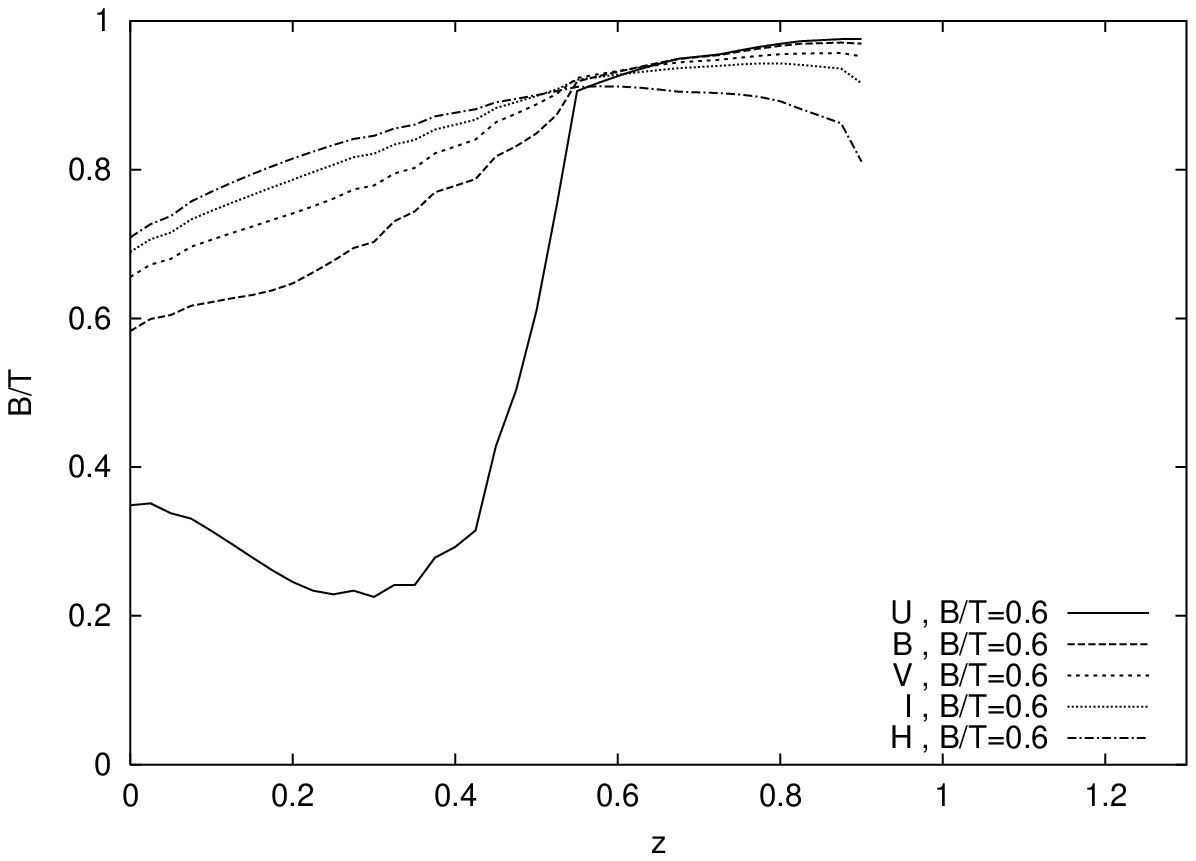}
\caption{Redshift evolution of B/T - ratios in various bands for two different 
galaxy types with present-day ${\rm (B/T)|_B = 0.1~and~0.6}$, respectively. Old 
disk and younger bulge.}
\label{fig:bt_m_od}
\end{figure}

In this scenario (\ref{fig:bt_m_od}) at redshifts higher than ${\rm z \sim 0.9}$, bulge SF did not
occur and only pure disks are expected.  At ${\rm z = 0}$, we recover the
results from Section \ref{sect:oo} showing the strong wavelength dependence of
the B/T ratio. Next, we note the tremendous differences in the redshift
evolution between the U-band B/T-ratios and those in longer wavelength
bands. For strongly bulge-dominated systems, the B/T-ratios in B, V, and I
evolve quite homogeneously in contrast to the previous scenario. They increase
slowly from ${\rm (B/T)|_B = 0.6}$ at ${\rm z = 0}$ to ${\rm B/T \sim 0.9}$ at
${\rm z \sim 0.8}$ in all bands except for U.

For galaxies with weak bulge contributions, e.g. ${\rm (B/T)|_B = 0.1}$ at ${\rm
z = 0}$, the B/T-ratios increase considerably in all bands B, V, I to ${\rm z
\sim 0.6}$, from whereon their slope has a turning point is beginning 
to decrease to ${\rm z \sim 0.9}$. Around ${\rm z \sim 0.6}$, the curves for
the B/T in different bands cross each other, i.e. at ${\rm z \gta 0.6}$ the B/T
values at shorter wavelengths become higher than those at longer ones in
contrast to the situation at lower redshift.

With I-band bulge-disk decomposition, the B/T ratio of a galaxy at ${\rm z \sim
0.5}$ would be overestimated as compared to local I-band B/T ratios by typically
a factor $\sim 2$ for low B/T systems and less so for higher B/T galaxies, except for the U-band.

In the I-band, the B/T ratio of a local Sc-type galaxy with
${\rm (B/T)|_B (z=0) = 0.1}$ increases from ${\rm (B/T)|_I=0.14}$ at ${\rm z=0}$ with increasing redshift to ${\rm
(B/T)|_I = 0.2}$ at ${\rm z \sim 0.2}$ and ${\rm (B/T)|_I = 0.5}$ at ${\rm z
\sim 0.8}$. This is due to evolutionary and cosmological effects. At ${\rm z
\sim 0.8}$, I-band observations sample rest-frame B where the bulge component at
an age of the disk of $\sim 5$ Gyr is actively forming stars in this scenario.

Hence, the intrinsically relatively low mass bulge components of later spiral
types are much more prominent in early evolutionary phases closer to their
active SF phase and cosmological corrections further enhance this evolutionary
effect.

Systems with ${\rm (B/T)|_I \sim 0.6}$ at ${\rm z \sim 0.5 - 0.9}$ are in fact
the progenitors of local late-type Sc galaxies with ${\rm (B/T)|_B = 0.1}$ at
${\rm z = 0}$.

\subsection{Model predictions for NICMOS observations}
Comparing in Figs. 4 - 6 the redshift evolution of Johnson's \citep{bb88} H-band B/T-ratios with those for the I-band discussed above, we see that both are fairly similar with the H-band evolution being slightly more continuous than that in I. The H-band hence appears slightly better suited for a comparison among galaxies over a range of redshifts than I. Note, however, that already at ${\rm z = 0}$ strong bulge systems with ${\rm (B/T)|_B = 0.6}$ show ${\rm (B/T)|_H \sim 0.75}$ and weak bulge systems ($\sim$ Sc) with ${\rm (B/T)|_B = 0.1}$ show ${\rm (B/T)|_H \sim 0.2}$. 

While for systems that are bulge-strong at ${\rm z = 0}$ the bulge components of their progenitors at higher redshift are slightly overestimated this effect is much stronger for the progenitors of local weak-bulge systems. H-band decomposition of the progenitors of local Sc galaxies are predicted by our models to yield B/T-light ratios of ${\rm (B/T)|_H = 0.4}$ around ${\rm z \sim 0.8}$ in case of equal age bulge and disk or for old disk and younger bulge components and as high as ${\rm (B/T)|_H = 0.7}$ for the old bulge and younger disk scenario. For the equally old bulge and disk scenario the H-band B/T-ratio of local weak-bulge ($\sim$ Sc) progenitor galaxies smoothly increases further at redshifts beyond 0.8 and levels off around ${\rm (B/T)|_H = 0.8}$ at ${\rm 2 \leq z \leq 4}$. 

\subsection{Cosmological and evolutionary corrections for bulge and disk components}
A quantitative comparison of bulge and disk decompositions over a range of redshifts -- unless restricted to I- or H-band investigations in a first approximation -- requires evolutionary as well as cosmological corrections that can be given by our models. 
 
For unresolved galaxies evolutionary and cosmological corrections are conventionally given in terms of magnitude differences, which are normalized to the average
luminosity of locally observed galaxies of the respective spectral types. Bulge and disk subcomponents, however,
do not have such an absolute magnitude scale. We therefore chose to give their
evolutionary and cosmological corrections in terms of luminosity ratios rather
than magnitude differences -- like in the case for single burst stellar populations (cf. Schulz \etal 2002) -- and call them ${\rm \epsilon_{\lambda}}$ and $
{\rm \kappa_{\lambda}}$, respectively, not to be confused with the ${\rm e_{\lambda}}$
and ${\rm k_{\lambda}}$ conventionally given for galaxies. With the luminosities ${\rm L_{\lambda}}$ for the bulge and disk components chosen as to match ${\rm L_{\lambda}^{gal}=L_{\lambda}^{bulge}+L_{\lambda}^{disk}}$ and ${\rm B/T|_{\lambda}=L_{\lambda}^{bulge}/(L_{\lambda}^{bulge}+L_{\lambda}^{disk})}$ the ${\rm \epsilon_{\lambda}}$ and ${\rm \kappa_{\lambda}}$ for bulge and disk, respectively, are defined as

\begin{eqnarray*}
{\rm \epsilon_\lambda := {L_\lambda (z, t(z)) \over L_\lambda(z,t_0)}} \\
{\rm \kappa_\lambda := {L_\lambda (z,t_0) \over L_\lambda (0,t_0)}},
\end{eqnarray*}
with ${\rm t(z)}$ being the age of a buge or disk observed at redshift ${\rm z}$ and ${\rm t_0}$ the age at ${\rm z=0}$. 

\begin{table}

\begin{tabular}[h]{cccccccc}
\hline
        ${\rm z}$ & ${\rm \epsilon_U}$ & ${\rm \kappa_U}$ & ${\rm \epsilon_B}$ & ${\rm \kappa_B}$ & ${\rm \epsilon_V}$ &...\\
\hline
0.03 & 1.0828 & 0.8461 & 1.0511 & 0.9286 & 1.0404&\\
0.05 & 1.1416 & 0.7082 & 1.0826 & 0.8462 & 1.0638&\\
0.08 & 1.2474 & 0.5911 & 1.1300 & 0.7597 & 1.0962&\\
\vdots \\
\hline
\end{tabular}

\begin{tabular}[h]{cccccccc}
\hline
        ${\rm z}$ & ${\rm \epsilon_U}$ & ${\rm \kappa_U}$ & ${\rm \epsilon_B}$ & ${\rm \kappa_B}$ & ${\rm \epsilon_V}$ &...\\
\hline
0.03 & 1.1321 & 0.8562 & 1.0909 & 0.9346 & 1.0794&\\
0.05 & 1.2118 & 0.7265 & 1.1494 & 0.8599 & 1.1321&\\
0.08 & 1.3444 & 0.6151 & 1.2568 & 0.7816 & 1.2323&\\
\vdots \\
\hline
\end{tabular}

\begin{tabular}[h]{cccccccc}
\hline
        ${\rm z}$ & ${\rm \epsilon_U}$ & ${\rm \kappa_U}$ & ${\rm \epsilon_B}$ & ${\rm \kappa_B}$ & ${\rm \epsilon_V}$ &...\\
\hline
0.03 & 0.9973 & 0.9608 & 0.9901 & 0.9629 & 0.9850 &\\
0.05 & 0.9965 & 0.9272 & 0.9856 & 0.9169 & 0.9780 &\\
0.08 & 0.9954 & 0.9003 & 0.9791 & 0.8704 & 0.9671 &\\
\vdots \\
\hline
\end{tabular}

\begin{tabular}[h]{cccccccc}
\hline
        ${\rm z}$ & ${\rm \epsilon_U}$ & ${\rm \kappa_U}$ & ${\rm \epsilon_B}$ & ${\rm \kappa_B}$ & ${\rm \epsilon_V}$ &...\\
\hline
0.03 & 0.9945 & 0.9649 & 0.9824 & 0.9667 & 0.9741 &\\
0.05 & 0.9925 & 0.9350 & 0.9741 & 0.9247 & 0.9615 &\\
0.08 & 0.9897 & 0.9114 & 0.9621 & 0.8827 & 0.9425 &\\
\vdots \\
\hline
\end{tabular}

\caption{Evolutionary and cosmological corrections ${\rm \epsilon_{\lambda}}$ and
${\rm \kappa_{\lambda}}$, respectively, as a function of redshift ${\rm z}$ from ${\rm z=0}$ 
to ${\rm z=5}$ in the cosmology ${\rm H_0=65}$, ${\rm \Omega_0=0.10}$ for all bands ${\rm U,~B,~.~.~.~,~K}$. The first two tables describe old and younger bulge components, respectively. The third and fourth table are for old and younger disk components. The full table 
for all wavelengths bands is given in the 
electronic version, at CDS, and on our web page (http://www.uni-sw.gwdg.de/$^{\sim}$galev).}

\label{e-korr}
\end{table}

\begin{table}
\begin{center}
\begin{tabular}[h]{ccc}
\hline
	z &  t [Gyr] & BDM \\
\hline 
	0.0 & $11.92$ & 31.15\\
	0.5 & $6.957$ & 42.267\\
	1.0 & $4.535$ & 44.130\\
	1.5 & $3.115$ & 45.305\\
	2.0 & $2.189$ & 46.179\\
	2.5 & $1.542$ & 46.877\\
	3.0 & $1.065$ & 47.461\\
	3.5 & $0.702$ & 47.961\\
	4.0 & $0.417$ & 48.400\\
	4.5 & $0.188$ & 48.790\\
	5.0 & $0.0$ & 49.141\\
\hline
\end{tabular}
\end{center}
\caption{Evolution time t and BDM as a function of redshift z for ${\rm H_0=65}$, ${\rm \Omega_0=0.1}$, and ${\rm \Lambda_0=0}$.}
\end{table}

In Table 3 we give the evolutionary and cosmological corrections ${\rm \epsilon_{\lambda}}$ and ${\rm \kappa_{\lambda}}$ as a function of redshift z for broad band filters ${\rm \lambda = U,~B,~.~.~.~,~K}$, separately for our old and younger bulge and old and younger disk models in a cosmology with ${\rm H_0=65}$, ${\rm \Omega_0=0.10}$.

\begin{equation*}
{\rm \epsilon_{\lambda}^{gal}(z)~=~\epsilon_{\lambda}^{disk}(z) \cdot (1 - (B/T)|_{\lambda}(z))} 
\end{equation*}

\vspace{-0.3cm}
\begin{equation*}{\rm \hspace{2.cm} + \: \epsilon_{\lambda}^{bulge}(z) \cdot (B/T)|_{\lambda}(z)}
\end{equation*}

\vspace{-0.3cm}
\begin{equation*}
{\rm \kappa_{\lambda}^{gal}(z)~=~\kappa_{\lambda}^{disk}(z) \cdot (1 - (B/T)|_{\lambda}(z))}
\end{equation*}

\vspace{-0.3cm}
\begin{equation*}{\rm \hspace{2.cm} + \: \kappa_{\lambda}^{bulge}(z) \cdot (B/T)|_{\lambda}(z)}
\end{equation*}

In analogy to the case of unresolved galaxies where the apparent magnitude at some redshift z in a filter $\lambda$, ${\rm m_{\lambda}(z)}$, is obtained from the absolute magnitude ${\rm M_{\lambda}}$ of the respective galaxy at the present age and redshift ${\rm z=0}$, its evolutionary and cosmological corrections, and the bolometric distance modulous BDM(z) via 
$${\rm m_{\lambda}(z) = M_{\lambda}+e_{\lambda}(z)+k_{\lambda}(z)+BDM(z),}$$ the apparent magnitude ${\rm m_{\lambda}^{gal}(z)}$ of some galaxy at redshift z with a local bulge-to-total-light ratio ${\rm (B/T)|_{\lambda}}$ is obtained from its local present day luminosity ${\rm L_{\lambda}^{gal}}$ via 

$${\rm m_{\lambda}^{gal}(z) = -2.5~ Log (L_{\lambda}^{gal} \cdot \epsilon_{\lambda}^{gal}(z) \cdot \kappa_{\lambda}^{gal}(z))+BDM(z)}$$. 

Table 4 gives the evolution time t and the bolometric distance modulous BDM as a function of redshift for a cosmology with ${\rm H_0=65}$, ${\rm \Omega_0=0.1}$, ${\rm \Lambda_0=0}$.

\section{Conclusions and Outlook}

The longstanding controversy about the origin of the different galaxy components
bulge and disk, in particular about their respective formation epochs and
mechanisms, may soon be solved. High resolution imaging of faint sources, e.g.
with HST, opens up the possibility of surface brightness decomposition into
${\rm r^{1/4}-}$ bulge and exponential disk components for galaxies up to
considerable redshifts ${\rm z > 1}$. Applied to reasonably large samples of
galaxies with observational biases carefully taken into account, this technique
will allow to directly study the redshift evolution of both components to
significant look-back times. To derive the redshift evolution of bulge and disk
luminosity contributions and study how far back in time the locally well
established correlation between galaxy morphological and spectrophotometric
properties will hold, however, careful consideration of both cosmological
band-shift and evolutionary effects are of crucial importance.

We present a first and simplified evolutionary synthesis approach to analyze
these two effects. Bulge star formation -- like that for any kind of spheroidal
system -- is conventionally assumed to occur on short timescales ($10^8 - 10^9$
yr) while the star formation rate in the disk component is safely assumed to be
roughly constant in time.

We separately model the spectral evolution from UV through NIR of both
components and superpose them in various proportions as to match by an age of 12
Gyr the average bulge-to-total light ratios as conventionally measured in the
B-band for different galaxy types (S0, Sa, . . ., Sd).  To cope with the full
range of currently discussed formations scenarios for both galaxy components, we
present models in which both bulge and disk star formation is assumed to start
around the same time in the early Universe (at ${\rm z \sim 5}$) and others
in which either of the two is somehow delayed until ${\rm z \sim 1}$.

We can then study in each of the three scenarios the wavelength dependence of
present-day bulge-to-total light ratios that results from the different star
formation timescales (and possibly ages) of the two components. While it is
intuitively clear that galaxy morphologies depend on the wavelength of
observation, we can show quantitatively the importance of this effect for the
decomposition of surface brightness profiles. It is strongest for weak bulge
systems and only slightly dependent on the adopted scenario. For an Sc-type
galaxy with B-band ${\rm B/T = 0.1}$ e.g., the B/T ratio increases by a factor 4
(old disk and younger bulge) to 6 (old bulge and old or younger disk) from U-
through I-band. A very first comparison shows recent observational results for a sample of galaxies decomposed both in B- and H-bands by \citet{esk} to be well consistent with our model predictions for I-band decomposition. The large scatter in the observed ${\rm (B/T)|_H}$ vs. ${\rm (B/T)|_B}$ relation, however, precludes us from discriminating between our three different scenarios. 

In a second step, we study for a standard cosmological model (${\rm H_0 = 65,~
  \Omega_0 = 0.1,~ \Lambda_0 = 0}$) the redshift evolution of bulge-to-total
light ratios and find it to be significantly different for the three different
scenarios. The progenitors of local galaxies of a given type, i.e., of a given
B-band B/T light ratio at ${\rm z = 0}$, are shown to feature strongly different
B/T ratios when observed over a range in redshift in the same band. The redshift
evolution of B/T ratios in a given band depends in a complex way on galaxy type,
i.e. present-day B/T ratio, on the wavelength band in which the profile
decomposition is performed, and on the formation epochs of the two components.

In particular, we find that in the scenario with old bulge and equally old disk
the progenitors of local weak bulge Sc-type galaxies are expected to show
significantly stronger bulge components at all wavelengths at ${\rm z \gta
2}$. This is also significant in the scenario with old disk and younger bulge at
${\rm z \gta 0.5}$.

In case of a younger bulge its strong and relatively recent star formation 
makes even low present-day B/T systems appear bulge-dominated at ${\rm z \gta 0.5}$.

The I- and H-bands generally show the smoothest redshift evolution in terms of
B/T light ratios and hence are the best wavelengths to compare B/T-ratios over some range in redshifts. Nevertheless, it is important to take into account that the progenitors of any type of local galaxy  with ${\rm B/T \geq
  0.1}$ are expected to show stronger bulge contributions at higher redshifts. 
So, even in the I- and H-bands, and much more so in other wavelength bands, B/T ratios
of galaxies at different redshifts can only be compared if evolutionary and
cosmological corrections are properly accounted for. 

Irrespective of the respective ages of the bulge and disk stellar
components, our models indicate that both for low and high present-day
B/T-galaxies, their I-band B/T-ratios rise with increasing
redshift. I.e., I-band B/T-ratios of galaxies at ${\rm z > 0}$ should
overestimate their bulge light contributions as compared to local
galaxies. \citet{mar}, however, report a scarcity of bulge-strong systems at ${\rm z \gta 1}$ from their I-band decomposition of HDF-N galaxy profiles. In view of our results the scarcity of bulge-strong systems at high redshift should even be more dramatic than what they find. The same holds true for the results recently reported by \citet{ag2002}  from an I-band B/T-decomposition of galaxies from two Hawaiian Deep Fields. The $\sim 30 \%$ decrease in the number density of E/S0 galaxies to ${\rm z = 0.8}$ they find should also be seen as a lower limit to their real decrease in view of our results. We recall that while surface brightness dimming may cause a selection bias against disks at high redshift this is not expected for the high surface brightness bulge components. 
 
We show that surface brightness profile decomposition in more than one band and
for a sample of galaxies at different redshifts (see \citet{simard02} for a first data set) should allow to discriminate
between the different scenarios for bulge and disk formation. It will also be useful to eliminate the effects of dust reddening and absorption shown to be very important for the proper bulge-disk decomposition of local galaxies by \citet{bp94}. 

We give electronic tables for the luminosity evolution of bulge and disk subcomponents as well as for their respective evolutionary and cosmological corrections for a standard csmological model in all scenarios considered. They allow for a detailed quantitative comparison of galaxy B/T ratios in various bands over a wide redshift range.  

Clearly, more observations, refined models, and, in particular, profile decompositions at more than one wavelength are required to definitively settle this important issue which has far-reaching consequences for our understanding of the formation and cosmological evolution of galaxies. 

We believe that as simplified as our first set of models is at the present
stage, they are a useful and indispensable tool for the interpretation of galaxy
morphologies at ${\rm z > 0}$ and their redshift evolution. Detailed comparison
with existing HST data (e.g. \citeauthor{mar} \citeyear{mar}, \citeauthor{ag2002} \citeyear{ag2002}, \citeauthor{simard02} \citeyear{simard02}) should allow to seriously
constrain galaxy formation scenarios.

We caution that all our conclusions result from the different star formation
timescales (and possibly different redshifts at which star formation sets in) in
the bulge and disk components. They are therefore not able to describe a
situation -- if it exists -- in which a bulge is formed without significant
additional star formation by a mere regrouping of disk stars due to dynamical
instabilities.


\begin{acknowledgements}
We thank our referee, Marc Balcells, for his valuable suggestions that helped to significantly improve the presentation of our results.
\end{acknowledgements}


\bibliography{paper2}

\begin{thebibliography}{38}
\expandafter\ifx\csname natexlab\endcsname\relax\def\natexlab#1{#1}\fi

\bibitem[{{Abraham} {et~al.}(1999){Abraham}, {Ellis}, {Fabian}, {Tanvir}, \&
  {Glazebrook}}]{1999MNRAS.303..641A}
{Abraham}, R.~G., {Ellis}, R.~S., {Fabian}, A.~C., {Tanvir}, N.~R., \&
  {Glazebrook}, K. 1999, MNRAS, 303, 641

\bibitem[{{Aguerri}(1999)}]{1999A&A...351...43A}
{Aguerri}, J.~A.~L. 1999, A\&A, 351, 43

\bibitem[{Aguerri \& Trujillo(2002)}]{ag2002}
Aguerri, J. A.~L. \& Trujillo, I. 2002, MNRAS

\bibitem[{{Andredakis} {et~al.}(1995){Andredakis}, {Peletier}, \&
  {Balcells}}]{andre1995}
{Andredakis}, Y.~C., {Peletier}, R.~F., \& {Balcells}, M. 1995, MNRAS, 275, 874

\bibitem[{{Balcells} \& {Peletier}(1994{\natexlab{a}})}]{1994AJ....107..135B}
{Balcells}, M. \& {Peletier}, R.~F. 1994{\natexlab{a}}, AJ, 107, 135

\bibitem[{{Balcells} \& {Peletier}(1994{\natexlab{b}})}]{bp94}
---. 1994{\natexlab{b}}, \aj, 107, 135

\bibitem[{{Bender} {et~al.}(1993){Bender}, {Burstein}, \&
  {Faber}}]{1993ApJ...411..153B}
{Bender}, R., {Burstein}, D., \& {Faber}, S.~M. 1993, ApJ, 411, 153

\bibitem[{{Bessell} \& {Brett}(1988)}]{bb88}
{Bessell}, M.~S. \& {Brett}, J.~M. 1988, PASP, 100, 1134

\bibitem[{{Bicker} {et~al.}(2002){Bicker}, {Fritze-v.~Alvensleben}, \&
  {Fricke}}]{bicker2002}
{Bicker}, J., {Fritze-v.~Alvensleben}, U., \& {Fricke}, K.~J. 2002, A\&A, 387,
  412

\bibitem[{Bruzual \& Charlot(1993)}]{bruz2}
Bruzual, G. \& Charlot, S. 1993, ApJ, 538

\bibitem[{{de Vaucouleurs}(1958)}]{1958ApJ...128..465D}
{de Vaucouleurs}, G. 1958, ApJ, 128, 465

\bibitem[{Eskridge {et~al.}(2002)Eskridge, Frogel, Pogge, Quillen, Berlind, \&
  Davies}]{esk}
Eskridge, P.~B., Frogel, J.~A., Pogge, R.~W., Quillen, A.~C., Berlind, A.~A.,
  \& Davies, R.~L. 2002, astro-ph/0206320

\bibitem[{{Fritze- v. Alvensleben} \& Gerhard(1994)}]{fritze1}
{Fritze- v. Alvensleben}, U. \& Gerhard, O. 1994, A\&A, 751

\bibitem[{{Guiderdoni} \& {Rocca-Volmerange}(1987)}]{1987A&A...186....1G}
{Guiderdoni}, B. \& {Rocca-Volmerange}, B. 1987, \aap, 186, 1

\bibitem[{{Hubble}(1926)}]{1926ApJ....64..321H}
{Hubble}, E.~P. 1926, ApJ, 64, 321

\bibitem[{{Jablonka} {et~al.}(1996){Jablonka}, {Martin}, \&
  {Arimoto}}]{1996AJ....112.1415J}
{Jablonka}, P., {Martin}, P., \& {Arimoto}, N. 1996, AJ, 112, 1415

\bibitem[{{Jones} {et~al.}(2000){Jones}, {Smail}, \& {Couch}}]{jo2000}
{Jones}, L., {Smail}, I., \& {Couch}, W.~J. 2000, ApJ, 528, 118

\bibitem[{{Kennicutt}(1998{\natexlab{a}})}]{1998ARA&A..36..189K}
{Kennicutt}, R.~C. 1998{\natexlab{a}}, \araa, 36, 189

\bibitem[{{Kennicutt}(1998{\natexlab{b}})}]{1998ApJ...498..541K}
---. 1998{\natexlab{b}}, ApJ, 498, 541

\bibitem[{{Kormendy}(1985)}]{1985ApJ...292L...9K}
{Kormendy}, J. 1985, ApJ, 292, L9

\bibitem[{{Maraston} \& {Thomas}(2000)}]{2000ApJ...541..126M}
{Maraston}, C. \& {Thomas}, D. 2000, ApJ, 541, 126

\bibitem[{Marleau \& Simard(1998)}]{mar}
Marleau, F. \& Simard, L. 1998, ApJ, 585

\bibitem[{{Martin} \& {Friedli}(1997)}]{1997A&A...326..449M}
{Martin}, P. \& {Friedli}, D. 1997, A\&A, 326, 449

\bibitem[{{McWilliam} \& {Rich}(1994)}]{1994ApJS...91..749M}
{McWilliam}, A. \& {Rich}, R.~M. 1994, ApJS, 91, 749

\bibitem[{{Moore} {et~al.}(1998){Moore}, {Lake}, \& {Katz}}]{mo1998}
{Moore}, B., {Lake}, G., \& {Katz}, N. 1998, ApJ, 495, 139

\bibitem[{{Norman} {et~al.}(1996){Norman}, {Sellwood}, \&
  {Hasan}}]{1996ApJ...462..114N}
{Norman}, C.~A., {Sellwood}, J.~A., \& {Hasan}, H. 1996, \apj, 462, 114

\bibitem[{{Peletier}(2001)}]{2001ApSSS.277..437P}
{Peletier}, R. 2001, Astrophysics and Space Science Supplement, 277, 437

\bibitem[{{Prugniel} {et~al.}(2001){Prugniel}, {Maubon}, \&
  {Simien}}]{2001A&A...366...68P}
{Prugniel}, P., {Maubon}, G., \& {Simien}, F. 2001, A\&A, 366, 68

\bibitem[{{Quilis} {et~al.}(2000){Quilis}, {Moore}, \& {Bower}}]{qu2000}
{Quilis}, V., {Moore}, B., \& {Bower}, R. 2000, Science, 288, 1617

\bibitem[{{Ram{\' i}rez} {et~al.}(2000){Ram{\' i}rez}, {Stephens}, {Frogel}, \&
  {DePoy}}]{2000AJ....120..833R}
{Ram{\' i}rez}, S.~V., {Stephens}, A.~W., {Frogel}, J.~A., \& {DePoy}, D.~L.
  2000, \aj, 120, 833

\bibitem[{{Rocha-Pinto} \& {Maciel}(1996)}]{1996MNRAS.279..447R}
{Rocha-Pinto}, H.~J. \& {Maciel}, W.~J. 1996, \mnras, 279, 447

\bibitem[{{Rocha-Pinto} \& {Maciel}(1998)}]{1998A&A...339..791R}
---. 1998, \aap, 339, 791

\bibitem[{{Sadler} {et~al.}(1996){Sadler}, {Rich}, \&
  {Terndrup}}]{1996AJ....112..171S}
{Sadler}, E.~M., {Rich}, R.~M., \& {Terndrup}, D.~M. 1996, AJ, 112, 171

\bibitem[{Schulz {et~al.}(2002)Schulz, v.~Alvensleben, Moeller, \&
  Fricke}]{schulz02}
Schulz, J., v.~Alvensleben, U.~F., Moeller, C., \& Fricke, K.~J. 2002, A\&A in
  press

\bibitem[{Simard {et~al.}(2002)Simard, Willmer, \& Vogt}]{simard02}
Simard, L., Willmer, C. N.~A., \& Vogt, N.~P. 2002, ApJS

\bibitem[{Simien \& de~Vaucouleurs(1986)}]{simien}
Simien, F. \& de~Vaucouleurs, G. 1986, ApJ, 564

\bibitem[{{Vazdekis}(1999)}]{1999ApJ...513..224V}
{Vazdekis}, A. 1999, ApJ, 513, 224

\bibitem[{{Wyse}(1999)}]{1999Ap&SS.267..145W}
{Wyse}, R.~F.~G. 1999, Ap\&SS, 267, 145

\end{thebibliography}
\bibliographystyle{apj}

\end{document}